\documentclass[aps,twocolumn,floatfix,,showpacs,superscriptaddress,footnoteinbib,prb]{revtex4-1}

\usepackage{amsmath, amsfonts, amssymb, mathrsfs, dsfont}
\usepackage{graphicx, color}
\usepackage{wasysym}
\usepackage{ulem} % needed for \scrap-command
\usepackage{bm}
\usepackage{dcolumn}   % needed for some tables
\usepackage[mathscr]{eucal}
\usepackage[dvipsnames]{xcolor}
\usepackage{mathtools}

\usepackage[colorlinks, breaklinks, 
            linkcolor=OrangeRed,
            citecolor=RoyalBlue,
            urlcolor=NavyBlue]{hyperref}
            
\usepackage[all]{hypcap} % let hyperlinks correctly point to figures rather than their captions; 

\newcommand{\be}{\begin{equation}}
\newcommand{\ee}{\end{equation}}
\newcommand{\bea}{\begin{eqnarray}}
\newcommand{\eea}{\end{eqnarray}}

\newcommand{\mfb}{\mathrm{b}}
\newcommand{\Kr}{\mathscr{K}}

\newcommand{\nn}{\nonumber}

\newcommand{\mc}{\mathcal}

\newcommand{\eq}[1]{\begin{align}#1\end{align}}

\newcommand{\dex}{{\Delta x}}

\newcommand{\Wc}{W_{\text{c}_1}}
\newcommand{\Wcc}{W_{\text{c}_2}}
\newcommand{\mblA}{{MBL$_\text{A}$} }
\newcommand{\mblB}{{MBL$_\text{B}$} }

\newcommand{\thop}{t_\text{h}} 
\renewcommand{\em}{\it} 
\begin{document} 
%\preprint{p29.C60Long}

\title{Slow dynamics and strong finite-size effects in many-body localization
with random and quasi-periodic potential}

\author{Felix Weiner} 
\author{Ferdinand Evers}
\affiliation{ Institute of Theoretical Physics, University of Regensburg, D-93040 Germany}
\author{Soumya Bera}
\affiliation{Department of Physics, Indian Institute of Technology Bombay, Mumbai 400076, India}

\date{\today}% It is always \today, today,

%\pacs{73.63.Rt, 73.63.-b, 73.23.Ad}% PACS, the Physics and Astronomy
%                             % Classification Scheme.
%Use showkeys class option if keyword
\begin{abstract}
%%%%%%%%%%%%%%%%%%%%%%%%%%%%%%%%%%%%%%%%%%%%%%%
We investigate charge relaxation in disordered and quasi-periodic
quantum-wires of spin-less fermions ($t{-}V$-model) 
at different inhomogeneity strength $W$ in the localized 
and nearly-localized regime. 
Our observable is the time-dependent density correlation function, 
$\Phi(x,t)$, at infinite temperature. 
We find that disordered and quasi-periodic models behave
qualitatively similar: 
Although even at longest observation times the width $\dex(t)$  of $\Phi(x,t)$
does not exceed significantly the non-interacting localization length,
$\xi_0$,  strong finite-size effects are encountered. 
Our findings appear difficult to reconcile with the 
 rare-region physics (Griffiths effects) that often 
 is invoked as an explanation for the slow dynamics 
observed by us and earlier computational studies.
%%
% As a relatively reliable indicator for the 
% boundary towards the many-body localized (MBL) regime even under these conditions,  
% we consider the exponent function $\beta(t) {=} d\ln \dex(t) / d\ln t$.  
%%
Motivated by our numerical data %  for $\beta$, 
we discuss a scenario in which the MBL-phase splits 
into two subphases: in \mblA  $\dex(t)$ diverges slower than any power, while it 
converges towards a finite value in \mblB. 
Within the scenario the transition between \mblA and the ergodic phase 
is characterized by a length scale, $\xi$, that exhibits an essential singularity 
$\ln \xi \sim 1/|W-\Wc|$. 
Relations to earlier numerics and proposals of two-phase scenarios will be discussed. 
\end{abstract}
\maketitle

%%%%%%%%%%%%%%%%%%%%%%%%%%%%%%%%%%%%%%%%%%%%%%%%%%%%%%%%%%%%%%%%%%%%%%%%%%%%%%%
\section{Introduction}
\begin{figure}[tbh]
  \includegraphics[scale=0.5]{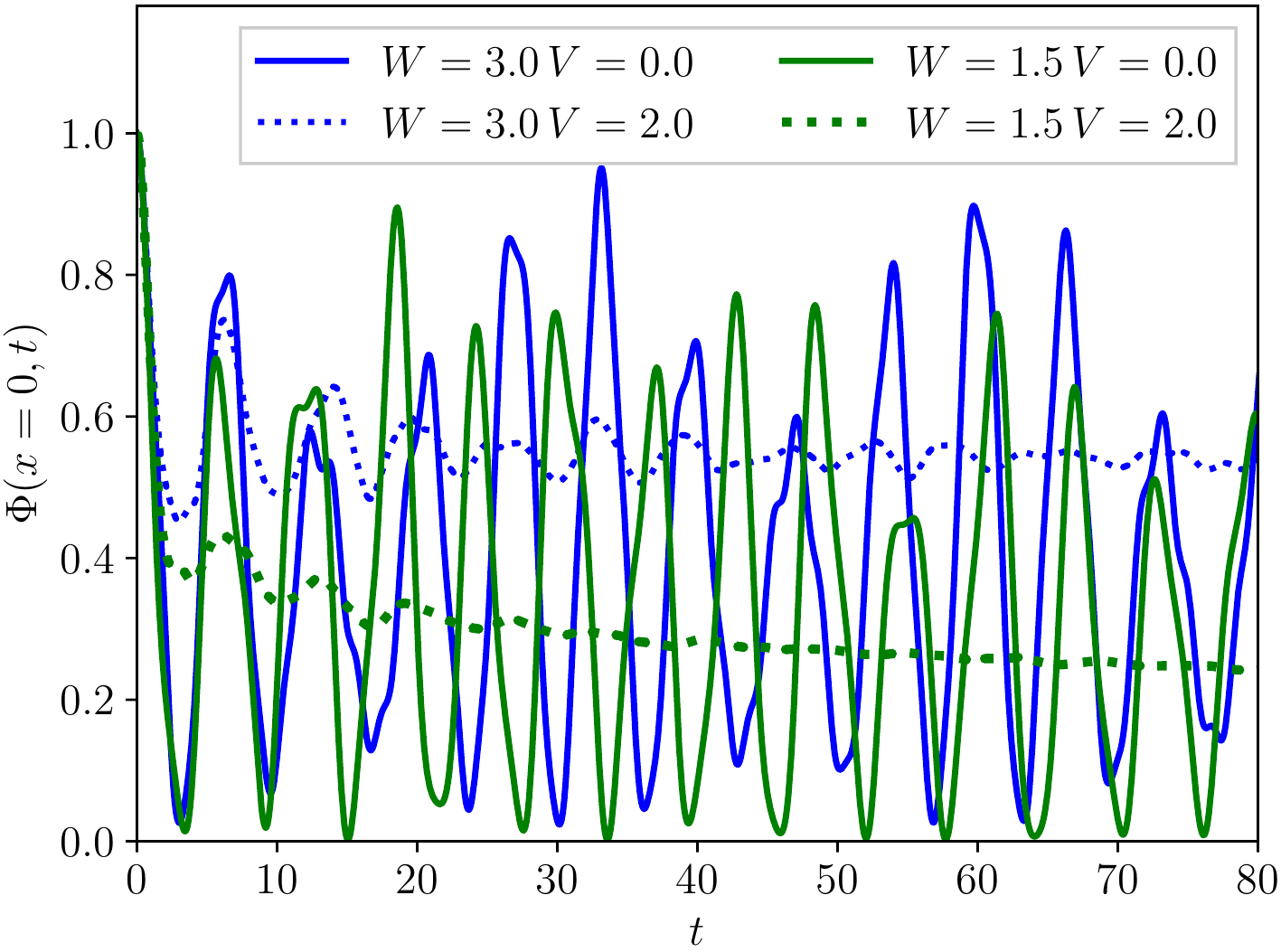}
  \caption{Return probability inside a disordered quantum wire
  ($t-V$-model) with and without two-body interactions (dashed vs solid traces) illustrating 
  the phenomenon of MBL. Two disorder strengths are shown situated 
  in the thermalizing phase (green, $W{=}1.5$) and closer to a possibly localized phase 
  (blue, $W{=}3.0$). While interactions wash-out the mesoscopic fluctuations very rapidly
  (strong dephasing), the time-averaged return probability, i.e. localization, 
  is affected only weakly (green) or hardly at all (blue). 
\label{f0}
% \label{Sf3}
} 
\end{figure}

%%%%%%%%%%%%%%%%%%%%%%%%%%%%%%%%%%%%%%%%%%%%%%%%%%%%%%%%%%%%%%%%%%%%%%%%%%%%%%%
%{\bf Introduction.} 
The interplay of quantum interference and interactions 
exhibits some of the most fascinating phenomena to be encountered 
in many-body systems. A typical representative of concern to 
us here is 
{\em many-body localization}~\cite{Basko2006,Gornyi2005}
 (MBL), which has attracted considerable attention over the last 
 decade~\cite{Znidaric2008, Pal2010, Bardarson2012, Nandkishore2015, Altman2015, Luitz2015, Bera2015, Znidaric2016, AnnMBLReview2017, Alet2018, AbaninBloch-Review-2018}. 
 At its heart is the basic observation that Anderson localization prevalent in low-dimensional, 
 disordered systems of non-interacting fermions can be robust against interaction effects at finite 
 temperature despite of dephasing as is illustrated in Fig.~\ref{f0}. 
%

%%%%%%%%%%%%%%%%%%%%%%%%%%%%%%%%%%%%%%%%%%%%%%%%%%%%%
%%%%%%%%%%%%%%%%%%%%%%%%%%%%%%%%%%%%%%%%%%%%%%%%%%%%%

Starting from the MBL phase, with decreasing disorder
a quantum-critical point is reached separating 
the MBL region from an ergodic one. 
Much progress has been made in recent years with respect to 
experimental investigations of the corresponding localization-delocalization transition
in diverse setups including cold atoms~\cite{Schreiber2015, Choi2016, Luschen2017, BordiaPRX2017, Kohlert2018, RispoliExp18}, 
%Rb optical lattice~\cite{RispoliExp18}, 
nuclear spin chains~\cite{WeiNuclearSpin18}, Yb ion system~\cite{Smith2016}, dipolar spins~\cite{KucskoDiamondPRL2018},  or in InO thin films~\cite{Ovadia2015}.
Importantly, in several experiments a very pronounced slow-down of the relaxation dynamics has
indeed been reported, when the (quasi-)disorder strength
exceeds a threshold value~\cite{Luschen2017, BordiaPRX2017, RispoliExp18}. 

The theoretical investigation of MBL and the related 
effects poses great challenges. On the one hand side 
the phenomenon escapes controlled analytical investigations. 
On the other hand, the computationally available window of observation times, $t_\text{obs}$, 
and system sizes often is too small in order to predict with 
confidence the thermodynamic, long time asymptotics. 
In fact as shown recently, due to dynamical slowing down,  
even in computational studies of 
the ergodic regime the width of the diffusion propagator, 
$\dex(t)$, is hardly ever seen to exceed the non-interacting 
localization length, $\xi_0$, significantly~\cite{Bera2017}. 
The resulting difficulties to safely extrapolate 
plague virtually all computational studies of 
dynamical behavior near and in the MBL phase. 
In fact, even present days experimental studies do not (yet) allow to go very deeply into the 
asymptotic regime.

It is noteworthy that in computations in 
addition to achieving sufficiently large system sizes and long observation 
times other difficulties arise;
we list two prominent examples: 
(i) Computer simulations carry nonphysical numerical parameters that need to be 
converged, such as the bond dimension, time increments and number of Chebyshev 
polynomials. 
Convergence can be difficult to achieve and will, in general, be sensitive to the 
calculated observable and the physical parameter regime. 
(ii) Quenches may produce transients or even asymptotic time traces 
that reflect properties of the initializing wavefunction more than the 
thermodynamic equilibrium. In this case the results of studies using different 
initialization, e.g. random versus staggered states, may be difficult 
to compare.\footnote{For example,~\textcite{Luitz2016}
report values for the imbalance exponent $\zeta$ 
that are smaller than the values detected by~\textcite{Doggen2018}
by a factor of two or more, e.g., near $W{=}2$. %, e.g. $\zeta\approx 0.08$ at $W\approx 2$.  
Both works have tested convergence with respect to numerical parameters, 
while they differ in the choice of the initial state: random~\cite{Luitz2016}
versus N\'eel~\cite{Doggen2018}. 
} 

Summarizing, a possibility to get stuck with transient or untypical 
behavior has to be accounted for. 
Nonwithstanding these difficulties, 
 from combining computational studies 
employing a variety of numerical techniques 
and analytical arguments based, e.g., on phenomenological renormalization-group treatments, 
a certain picture has emerged. It has been surveyed 
in several recent reviews~\cite{AbaninBloch-Review-2018,Vosk2015, Potter2015, Agarwal2017, Luitz:2017cp, Prelovsek-review-2017, Alet2018}. 

%%%%%%%%%%%%%%%%%%%%%%%%%%%%%%%%%%%%%%%%%%%%%%%%%
\begin{figure}[b]
\centering
 \includegraphics[width=1.0\columnwidth,keepaspectratio=true]{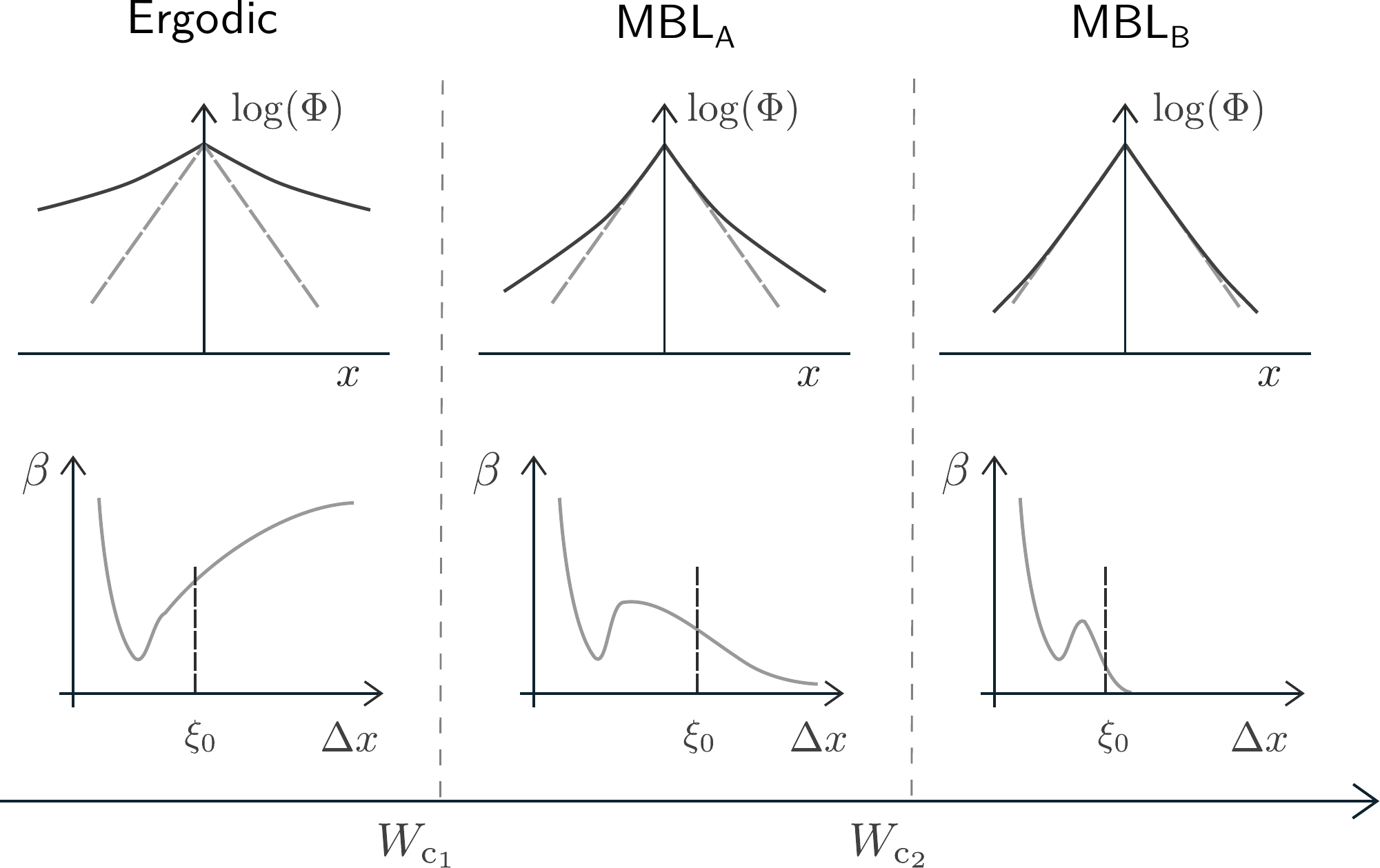}
\caption[MBL phase diagram]
{Proposed infinite temperature phase diagram of the $t{-}V$-model indicating the 
possible splitting of 
the MBL-phase into subphases. 
Upper panel: schematic behavior of density-density correlation function $\Phi(x,t)$. 
Lower panel: corresponding effective exponent $\beta$
(defined in Eq. \eqref{e1}) 
plotted over the variance of $\Phi(x,t)$, which is $\dex (t)$; 
$\xi_0$ indicates the non-interacting localization length. 
Between ergodic phase and \mblA, $\dot \beta(x\approx \xi_0)$ changes sign; 
once \mblB is reached, $\beta$ vanishes at a finite $\dex$. 
\label{f1}}
\end{figure}
%%%%%%%%%%%%%%%%%%%%%%%%%%%%%%%%%%%%%%%%%%%%%%%%%%%%%%
%%%%%%%%%%%%%%%%%%%%%%%%%%%%%%%%%%%%%%
%%%%%%%%%%%%%%%%%%%%%%%%%%%%%%%%%%%%%%%
In this work we present a computational study of the MBL region employing 
the $t{-}V$-model with fully random and quasi-periodic (Andre-Aubry-type) potentials~\cite{Iyer13, Lee17, KhemaniPRL17, Setiawan2017, Doggen2019}. We invoke as observable the 
density-density correlation function, $\Phi(x,t)$. 
As a first characterization its variance, $\Delta x(t)$, and the associated exponent function 
\begin{equation}
\label{e1} 
\beta(t) = \frac{d \ln  \Delta x(t)}{ d \ln t}. 
\end{equation} 
is being analyzed. 
Even though the diffusion propagator $\Phi(x,t)$ 
has been studied before,~\cite{BarLevPRL2015, BarlevEPL2016, Luitz:2017cp, Prelovsek-review-2017}
it is not the usual object of investigation.
When the exponent $\beta$ has been considered,  it has mostly 
been derived from closely related observables, such as the frequency dependent conductivity at zero wavenumber 
calculated in finite size systems~\cite{AgarwalPRL2015, GopalakrishnanPRB15, KhaitPRB16, SteinigewegPRB16, 
PrelovsekPRB16}. 
The density response at wavenumber $q{=}\pi$ (imbalance) and the 
corresponding dynamical exponent $\tilde \beta$ has been observed
more frequently~\cite{Luitz:2017cp, Doggen2018}, though 
its relation to $\beta$ is not well understood, in general. 
Also the exponent of the return probability, $\alpha$, 
has been employed, although the relation to $\beta$ involves an extra 
exponent, which is not known with good accuracy~\cite{Bera2017}.

\begin{figure}[b]
\centering
\includegraphics[width=1.0\columnwidth]{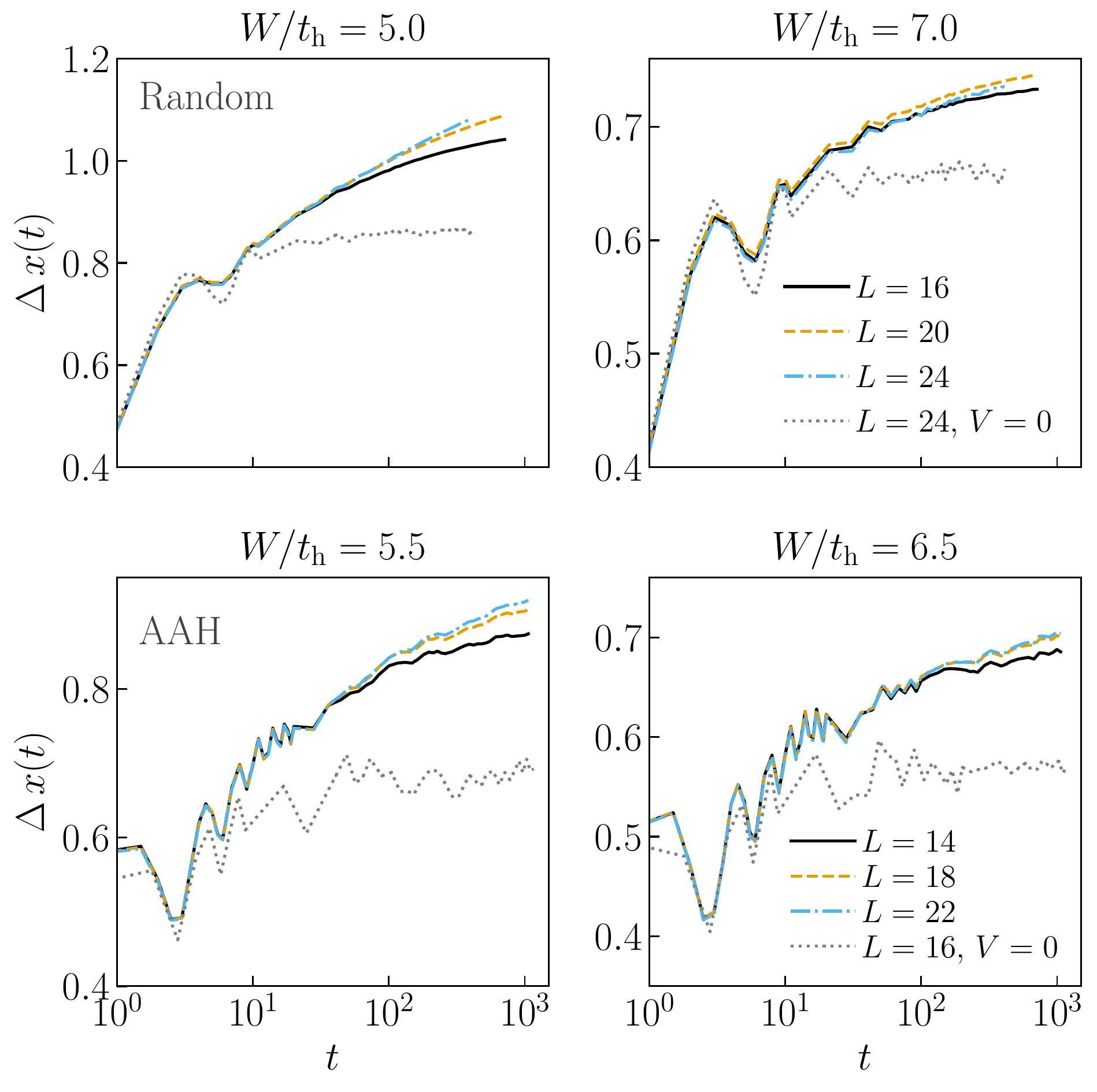}
\caption[Creeping variance]
{Time evolution of $\Delta x(t)$ with disordered (random, upper row) and quasi-periodic  (lower row) potential in \mblA (left) and closer to \mblB (right) phases as obtained in $t{-}V$ model.  In \mblA a very slow growth (``creep'') of $\Delta x(t)$ is observed in both potential-types; note that with increasing system size the slope at longest times keeps increasing, so that there is no evidence that the growth could be bounded. 
(Simulation parameters: 
$V/t_\text{h}=2.0$;  $W/t_\text{h}=\{5.0, 7.0\}$ (random) and $W/t_\text{h}=\{5.5, 6.5\}$ (quasi-periodic).)
\label{f2}}
\end{figure}

The interest of previous authors has been in the long time limit, 
$\beta_\infty(W) {\equiv} \beta(t{\to}\infty)$, 
or closely related exponents, where $\beta$ can serve as 
an effective ``order parameter'';
in the ergodic region prevailing at disorder $W{<}W_\text{c}$ one has 
$\beta_\infty{>}0$, while the MBL-phase, $W{>}W_\text{c}$, 
has been defined by $\beta_\infty{=}0$. 
% \SB{This footnote contains a lot of reference thus I think should be in the text. 
%{\color{green} Logical breech - let us discuss}}
% 
However, the data we here report 
leaves open a possibility for an intermediate window, $\Wc{<}W{<}\Wcc$,  
which constitutes within the MBL-phase a subphase \mblA; 
It exhibits a width $\Delta x(t)$ diverging 
in time but with a growth weaker than any power, 
see Fig.~\ref{f1} for a sketch and Fig.~\ref{f2} for illustrating data. 
At very large disorder, $W{>}\Wcc$, one enters the 
subphase, \mblB, in which $\beta(t){\to} 0$ upon $\dex(t)$ approaching the localization length $\xi(W)$. 
%%%%%%%%%%%%%%%%%%%%%%%%%%%%%%%%%%%%%%%%%%%%%%%%%
%%%%%%%%%%%%%%%%%%%%%%%%%%%%%%%%%%%%%%%%%%%%%%%%%

In view of the difficulty accessing the asymptotic regime of $\beta(t)$, 
we consider as a pragmatic indicator in addition to $\beta$ also 
its slope, $\dot \beta$, taken at the computationally still accessible
(i.e. relatively short) times $t_\xi$, 
where  $\Delta x(t)$ has passed the largest relevant microscopic length. 
\footnote{We include $\dot \beta(t)$ as an indicator with the idea that
$t_\xi$ is the only relevant time scale. Correspondingly, 
at $t\gtrsim t_\xi$ the sign of $\beta$ is converged, 
while it numerical value may not be, yet.}
Following this idea, entering the \mblA-phase from the ergodic side at $W{=}\Wc$
is signalized by a change of sign in $\dot \beta(t)$ at $t\gtrsim t_\xi$. While 
the ergodic phase is characterized by $\beta(t){>}0, \dot \beta(t){>}0$, 
we have $\beta(t){>}0,\dot \beta(t){<}0$ in {\mblA}. 
A  natural candidate for the microscopic length appearing in $t_\xi$ 
is the localization length, $\xi_0(W)$, of the non-interacting reference system, 
possibly having undergone an 
interaction-mediated (static) renormalization. Hence, we define $t_\xi$ implicitly via 
$
\label{e2} 
\Delta x(t_\xi) \approx \xi_0. 
$

%%%%%%%%%%%%%%%%%%%%%%%%%%%%%%%%%%%%%%
\begin{figure*}[t]
\centering
\includegraphics[width=1.0\textwidth]{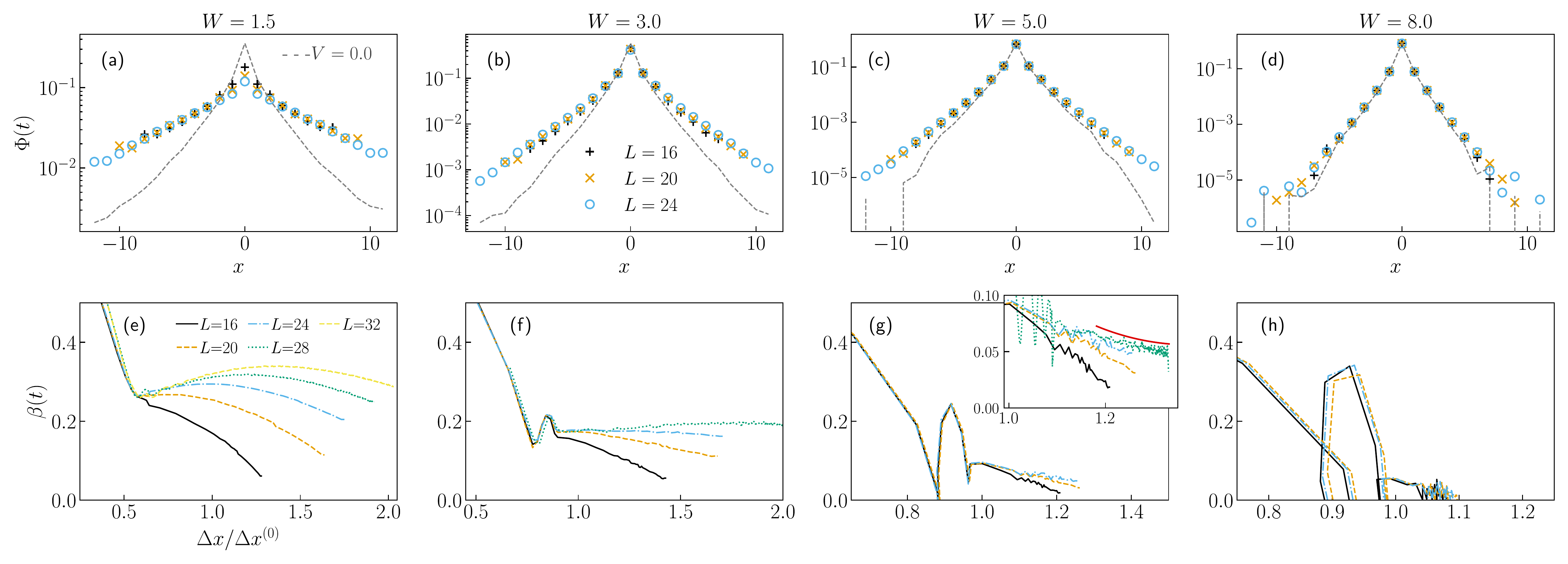}
\caption[Creeping variance and associated $\beta$-function]
{Effect of interactions with increasing disorder monitored by the density-density 
correlation function $\Phi(x,t)$; the analogous plot for the quasi-periodic model is 
shown in appendix. %~\cite{SuppMat}. 
Upper row:  $\Phi(x,t)$ for system sizes 
$L{=}16,20,24$
%$L{=}16,20,24,26$ 
and disorder $W{=}1.5,3.0,5.0,8.0$ at $t{\approx}400t^{-1}_\text{h}$. 
The non-interacting result at $L{=}24$ is also shown for reference (dashed line). 
The plot highlights how creep begins to manifest at larger disorder by ``lifting the tails''
of $\Phi(x,t)$. 
Lower row: The exponent function $\beta(t)$ corresponding to first-row data
plotted over $\dex(t)/\dex^{(0)}$
as suggested by Eq.~\eqref{e3}; 
 $\dex^{(0)}$ denotes the saturation width of the non-interacting diffusion propagator 
 at long time, here taken as an indicator of $\xi_0$.  
 $\beta$ inherits finite-size corrections from $\Phi(x,t)$, which persists also at larger 
 disorder values as is seen, e.g., in the inset. The red line in there serves as guide to the eye. 
The evolution seen in this data condenses into the schematics displayed in Fig.~\ref{f1}.
The noise visible at larger $W-$values reflects statistical noise from disorder averaging.
The average was taken over $\sim 10000$ samples for largest system size and for largest disorder values, while for smaller system sizes~($L=16, 20$) the disorder averaging is performed over $\sim 20000$ samples. 
\label{f3}}
\end{figure*}
% %%%%%%%%%%%%%%%%%%%%%%%%%%%%%%%%%%%%%

% %%%%%%%%%%%%%%%%%%%%%%%%%%%%%%%%%%%%%
We list the salient aspects of our numerics and the scenario displayed  
schematically  in Fig.~\ref{f1}: 
(i) In the ergodic phase, 
$\Delta x(t) \sim t^{\beta}$, where 
$\beta(t)$  converges to a  constant 
with $\dot \beta(t)$ approaching  zero from above; 
correspondingly, the slope $\dot \beta(t_\xi)$  seen in the numerical 
data should be positive when converged with respect to computational parameters.  
In \mblA the asymptotic growth of $\Delta x(t)$ is also
unbound, but $\beta(t){\to}0$ with $\dot \beta(t)$ approaching zero from below; 
correspondingly, $\dot \beta(t_\xi)$ seen in the numerical 
data is negative when converged.  
Finally, in \mblB  $\Delta x(t)$ is bounded from above 
and $\beta$ extrapolates to zero at a finite $\dex$. 
We interpret the numerical data shown in Fig. \ref{f3} as reflecting 
the evolution with increasing disorder strength $W$
described here and in Fig. \ref{f1}. 
The existence of an intermediate phase has been proposed 
before\cite{Dumitrescu2018}; the relation to our scenario 
is discussed below. 

(ii) We have applied our analysis method to the $t{-}V$ model with fully 
disordered and quasi-periodic potential. Importantly, comparing both cases we cannot detect an essential difference in the evolution of $\dex(t)$ with the inhomogeneity strength. Therefore, our results seem hard to reconcile 
with Griffiths-type scenarios that rely on rare-regions to explain  
the origin of slow dynamics~\cite{Vosk2015,Agarwal2017, Luitz:2017cp, Bar_Lev_2017}. 

(iii) To formulate our scenario  
we adopt the spirit of mode-coupling theory  
extrapolating into the asymptotic regime.
It suggests that the correlation length $\xi$, 
when entering \mblA from the ergodic phase, 
exhibits an essential singularity rather than a power-law, 
which is consistent with the Harris-Chayes criterion~\cite{Chayes1986}.
%%%

\section{Model and Computational Method} 
% {\bf Model and computational method.} 
 % %%%%%%%%%%%%%%%%%%%%%%%%%%%%%%%%%%%%%
We consider the $t{-}V$-model  
%%%%%%%%%%%%%%
\eq{
\mc{\hat{H}}  = & -\frac{t_\text{h}}{2}\sum_{x=1}^{L-1} \hat{c}^\dagger_{x} \hat{c}_{x+1} + h.c.  + 
\sum_{x=1}^{L}  \epsilon_x  \left ( \hat{n}_x - \frac{1}{2}\right) \nn \\
 &  + V\sum_{x=1}^{L-1}   \left ( \hat{n}_x - \frac{1}{2}\right)   \left ( \hat{n}_{x+1} - 
\frac{1}{2}\right), 
\label{e4}
}
%%%%%%%%%%%%%%
where hopping is over  $L$ sites with amplitude~$t_\text{h}$. 
We work at a half filling and with hard-wall  boundary conditions.
When employing disordered potentials, the onsite-energies, $\epsilon_x$, 
are  drawn from a box distribution $\lbrack-W,W\rbrack$.
\footnote{\label{fn23}
When working with quasi-periodic potential 
(Aubry-Andr\'e model)  we choose, 
$\epsilon_x {=} W \cos(2\pi \alpha x + \phi)$ with $\alpha=2/(\sqrt{5}+1)$
and $\phi$ is the random phase distributed uniformly between $[0, 2 \pi]$}
In subsequent calculations the interaction is taken 
twice larger, $V{=}2t_\text{h}$, than in previous studies~\cite{Bera2017}
to emphasize interaction effects. At the disorder values shown 
in Fig. \ref{f2}, the spectral statistics is close to Poissonian 
and in this sense the system is many-body localized, 
see Sec. \ref{si1}. 
% supplemental information~\cite{SuppMat}. 

Our observable is the density-density correlator, 
\begin{eqnarray}
\label{e5}
 \Phi(x,t)
 &=& 
 \overline{\langle (\hat n_x(t)-\frac{1}{2}) (\hat n_0-\frac{1}{2})\rangle} \Theta(t), 
% \label{e9}
\end{eqnarray}
with primary focus on the second moment, 
$
\Delta x(t)= \left[ \sum_{x=-L/2}^{L/2} x^2 \Phi(x,t)  - (\sum_{x=-L/2}^{L/2} x \Phi(x,t) )^2\right]^{1/2}. 
\nonumber
$
The disorder average is indicated by an overline; 
the thermal average is performed at infinite temperature and 
denoted by angular brackets, $\langle \ldots \rangle$; 
corresponding traces are performed stochastically. 
\footnote{Here, we have used that in the infinite temperature 
limit the density is homogeneous and equal to $1/2$.} 
For the time evolution, Eq.~\eqref{e5}, we employ
a standard Chebyshev-polynomial propagation~\cite{Wei06};
further details are given in Ref.~\onlinecite{Bera2017}.

% %%%%%%%%%%%%%%%%%%%%%%%%%%%%%%%%%%%%%

% %%%%%%%%%%%%%%%%%%%%%%%%%%%%%%%%%%%%%
\section{Results. \label{s.Results}} 
\subsection{Simulations} 
% {\bf Results. \label{s.Results}} 
% %%%%%%%%%%%%%%%%%%%%%%%%%%%%%%%%%%%%%
We systematically analyze traces of the kind shown in Fig.~\ref{f2} 
in terms of the logarithmic derivative \eqref{e1}; 
the results are shown in Fig.~\ref{f3}. 
The top-row displays the evolution of $\Phi(x,t)$ with system size and disorder
at $t{\mathrel{\mathop:}=}t_\text{obs}{\sim} 400 t^{-1}_\text{h}$. 
Fully consistent with our earlier findings, 
$\Phi(x,t)$ exhibits a pronounced non-Gaussian, nearly exponential shape 
-- here seen in a wide range of $W$-values. 
Not surprisingly, the interaction effects on $\Phi(x,t)$ are 
strongest in the tails; they emerge gradually, 
with ever increasing system size $L$ and observation time. 
Note that at larger disorder the interacting and non-interacting traces
are seen to almost coincide for smaller distances at the time observed, 
see, e.g., Fig.~\ref{f3}(d), $W{>}5$. At these 
(possibly transient) times one has $\dex(t)\approx \dex^{(0)}$. 
Therefore, interaction-related static renormalization effects of the 
disorder-strength are small at $W\gtrsim V$. 

The lower row of Fig.~\ref{f3}(e-h) is the basis of the schematic plot Fig.~\ref{f1}. 
It shows the exponent function $\beta(t)$ that highlights the important effects 
related to finite $L$. 
Consistent with our previous findings~\cite{Bera2017},
at weak disorder $W{\lesssim}3$ and $\dex(t){\sim}\dex_0$ there is a system size 
convergence with respect to the sign of the slope of $\beta(t)$; there is no 
notion of convergence at $\dex(t){\gtrsim}\dex_0$ beyond the statement that 
$\beta(\dex)$ is curved to the right. 
At larger disorder, $W{\gtrsim}5$, a qualitative change is observed: 
the curvature seen in the inset of Fig.~\ref{f3}(g) is about to turn left 
at $\dex(t){>}\dex_0$. 
The \mblA scenario advocated in Fig.~\ref{f1} is based on the observation that the 
curvature of $\beta(\dex)$ changes sign with growing $W$, 
so what is a right-curvature in Fig.~\ref{f3} turns into a left-curvature at later times; 
also we assume that there is no turn-around in $\beta(\dex)$ at even longer times, 
but rather a smooth decay towards the horizontal axis at $\dex{\to}\infty$. 
(A discussion of the pronounced short-time feature of the data Fig.~\ref{f3}(e-h) lower row, 
namely the spike growing in $\beta(\dex)$ with increasing $W$ near $\dex_0$, 
is given in the appendix.) 
%supplemental information~\cite{SuppMat}.)

%%%%%%%%%%%%%%%%%%%%%%%%%%%%%%%%%%%%%%%%%%%%%%%%%
% \section{Discussion\label{s.Discussion}} 
\subsection{Phenomenological modeling. \label{s.PM}}
% {\bf Phenomenological modeling. \label{s.PM}} 
% %%%%%%%%%%%%%%%%%%%%%%%%%%%%%%%%%%%%%
The data for the $\beta$-function displayed in Fig. \ref{f3} condenses 
into the phenomenological ansatz 
 \be
 \label{e3} 
 \beta[\Delta x] = \frac{\mfb}{1+\mfb_1 \Kr (\xi_0/\Delta x)}. 
 \ee
Similar to our previous work~\cite{Bera2017}, 
we here consider $\beta$ as a functional of the 
 slowest dynamical mode, %~\cite{BrenigII}, 
 which for simplicity we parametrize with $\dex(t)$; 
the parameters $\mfb,\mfb_1$ and the kernel $\Kr$ are evolving with disorder $W$. 
The ansatz helps formulating a scenario for 
the qualitative properties of the long-time limit, 
which are just barely seen to emerge in Fig. \ref{f3}.
% \sbc{,hardly visible irrespective of the disorder strength.} 
%

%%%%%%%%%%%%%%%%%%%%%%%%%%%%
Embarking on \eqref{e3} we  describe a  passage 
from the ergodic regime through \mblA 
into \mblB with increasing $W$.
The  kernel $\Kr(y)$ % displayed in \eqref{e3} 
equals unity at $y{=}1$ 
and otherwise is assumed to be 
a monotonous function in the interval 
$1{\gtrsim} y {\gtrsim} 0$. 
Its precise form is evolving with the 
disorder strength, $W$: With decreasing $y$, the kernel decreases in
the ergodic phase, increases in \mblA and is seen to diverge at a finite $y$ 
upon entering \mblB.

{\it Ergodic regime $W{<}\Wc$.}
The power-law growth of $\Delta x(t)$ characteristic for the ergodic 
regime is captured by  
\eqref{e3} with $\mfb{>}0, \mfb_1{>}0$ and taking $\Kr(y)$ 
as decreasing to zero in the limit $y\to 0$. 
Namely,  with the variance $\Delta x(t)$ growing in time, 
the kernel $\Kr(\xi_0/\Delta x(t))$ 
vanishes and so the denominator converges to unity. 
At this point, $\mfb$ should be identified with the diffusion exponent: 
$\mfb=\beta_\infty$

{\it Transition into \mblA.}
The transition is associated with a vanishing slope, 
$\dot \beta{=}0$. 
Within in the setting \eqref{e3} this requires the existence of a particular 
value of the disorder $\Wc$, such that the contribution of the term containing  
$\mfb_1 \Kr(\lambda/\Delta x)$ is independent of $\Delta x$. 
This situation could be realized in several ways; 
the most plausible scenario % within the framework \eqref{e3} 
is $\left. \Kr(y)\right|_{\Wc}=1$.
%%%%%%%%%%%%%%%%%%%%%%%%%%%%%%%%%%%%%%%%%%%%%%%%%%%%%%%%%%%%%%%%%%%%%%%%%%%%%%%%%%%%%%%5
\footnote{ 
Alternative possibilities could be 
(i)  $\mfb_1(\Wc){=}0$ or (ii) $\mfb_1(\Wc) {\to}\infty$. 
However, in our scenario we do not follow these directions.
% 
% The case (i) can be discussed when assuming the simplest possibility,
% which is that $\mfb_1(W)$ crosses zero at $\Wc$.  
% In this case, the sign change associated with the crossing 
% would imply according to \eqref{e3}  
% that transport continues to be subdiffusive once $\mfb_1$ has changed sign;  
% the effective diffusion exponent $\beta(t)$ then would be  
% decreasing in time. This behaviour is not inconsistent with 
% the numerical data, see Fig. \ref{f3}, e.g., at $W{=}5$.
% Nevertheless, we do not follow this direction, here.
% Namely, a regime in disorder with $\mfb_1(W)$ negative implies 
% within the framework of \eqref{e3} 
% the possibility a pole at a finite value 
% of $\xi$, which can be avoided only by assuming 
% a kind of parametrical fine-tuning. 
% %
% The case (ii) is readily ruled out. 
% It implies that at the transition $\dot\beta{=}0$ and $\beta{=}0$
% vanish simultaneously. Also this behaviour is not consistent with the data Fig. \ref{f3}. 
}
%%%%%%%%%%%%%%%%%%%%%%%%%%%%%%%%%%%%%%%%%%%%%%%%%%%%%%%%%%%%%%%%%%%%%%%%%%%%%%%%%%%%%%%%%%%
We thus parameterize $\Kr(y)$ in the vicinity of $W\approx \Wc$ as follows:
\be
\label{e6}
\Kr(y) \approx \tilde \Kr(y)^{W-\Wc}, \quad \tilde \Kr(y) := e^{\left. \frac{d\ln \Kr(y)}{dW}\right|_{\Wc}}. 
\ee
This shape has consequences for the correlation lengths, 
$
 \label{e7}
 1 {\approx} \mfb_1\ \Kr(\xi_0/\xi).  
$
which we here define implicitly; we have
$
\tilde \Kr(\xi_0/\xi) \approx \mfb_1^{\frac{1}{(\Wc-W)}}.  
$
To illustrate the impact of this statement, we adopt a 
power-law shape  $\tilde \Kr(\xi_0/\xi)\sim (\xi/\xi_0)^\kappa$, $\kappa{>}0$, so that 
\be
\xi \sim \xi_0\mfb_1^{\frac{1}{\kappa(\Wc-W)}}.  
\ee
It is thus seen that the scenario formulated in \eqref{e3} 
tends to generate length scales that are diverging at the transition 
in an exponential rather than in a power-law fashion. 

%%%%%%%%%%%%%%%%%%%%%%%%%%%%%%
{\it \mblA-phase.} 
According to \eqref{e6}, in \mblA the kernel $\Kr(y)$ diverges with $y{\to}0$ 
keeping $\mfb{>}0$ and $\mfb_1{>}0$. With growing time, 
the second term in the denominator grows without bound and eventually 
$
\beta(\Delta x) \approx \frac{\mfb}{\mfb_1} \ \Kr \left(\xi_0/\Delta x \right)^{-1}.  
$
The behavior is consistent with the data Fig. \ref{f3} as seen, e.g., 
at $W{=}5$. It predicts for the $\beta$-function a decay
approaching zero upon $\dex(t){\to}\infty$.

%%%%%%%%%%%%%%%%%%%%%%%%%%%%%%%%%
{\it Transition into the \mblB-phase.} 
Upon approaching $\Wcc$ and \mblB the kernel $\Kr(y)$ diverges at a 
nonvanishing value $y_c$. For illustration we adopt the generic form 
$
\Kr(y) = \frac{y^{\alpha}}{1+\mfb_2 y^{1/\nu}}
$
with $\alpha, \nu>0$. 
In \mblA, we have $\mfb_2{>}0$ and so the kernel 
diverges in a power-law fashion, $\Kr(y)\sim y^{\alpha-1/\nu}$, as required. 
When $W$ crosses $\Wcc$ from below, 
the coefficient $\mfb_2(W)$ changes sign from 
positive to negative, so in \mblB the kernel diverges at 
$
y^{1/\nu}=|\mfb_2|^{-1}. 
$
we conclude that 
the localization length diverges 
 $\xi\approx \lambda |\Wcc-W|^{-\nu}$.

%%%%%%%%%%%%%%%%%%%%%%%%%%%%%%%%%%
\subsection{Discussion} 
%{\bf Discussion.} 

The model \eqref{e3} has implications, 
which can be tested further. We mention the following two: 
(i) When entering \mblA from the ergodic side, 
$\beta_\infty(\Wc)$ is taking a non-vanishing, critical value.
In contrast, at the critical point separating \mblB and \mblA
$\beta_\infty(\Wcc){=}0$ is expected.
(ii) Previous computational studies have attempted a scaling analysis 
near $W{\approx}\Wc$ in order 
to extract a localization length exponent based on a power-law 
dependency $\xi{\sim}|W-\Wc|^{-\nu}$. 
The value of $\nu$ has not consolidated, yet: 
\textcite{Luitz2015} obtained $\nu{\approx}0.7$ 
from the system size scaling of the magnetization density, 
violating the Harris-Chayes bound $\nu{>}1$ in one 
dimension~\cite{Chayes1986}. 
This led to a subsequent discussion~\cite{Khemani2017}
attempting at rationalizing this result. 
On the other hand, 
recent exact diagonalization studies of the scaling of the  
Schmidt gap~\cite{Gray2018} or local temperature fluctuations~\cite{Lenarcic2018} found exponents exceeding $2$, which is consistent with the Harris bound.
We point out that within the scenario here lined out 
the transition that has been investigated in these
studies would corresponds to the entry into \mblA, 
so the correlation length diverges stronger than any power.
Hence, the Harris-Chayes bound is satisfied automatically. 

Saying this we emphasize once again the 
exploratory character of the scenario developed with 
Eq.~\eqref{e3}. Due to creep, 
we do not see at present a reliable way of telling, 
whether the system actually localizes or not. 
We stress that creep is a phenomenon that gradually emerges at largest system sizes, 
longest time scales and only in the tails of the distribution \eqref{f3}.
For observing or ruling out creep with frequently employed observables, such as the 
level statistics, the return probability or the sublattice imbalance, deviations have to be 
identified from the non-interacting background-behavior; such deviations   
manifest as small signals towards delocalization that slowly grow with 
increasing time and system size and thus are challenging to consolidate. 
% ~\cite{SuppMat}. 
% We relegate further discussion about relation of our results to earlier 
% RG studies and other numerical works to supplemental material.

%%%%%%%%%%%%%%%
\section{Relation to earlier studies}
\subsection{RG-studies.} 
%%%%%%%%%%%%%%%
We compare our scenario with recent renormalization group (RG) 
calculations~\cite{Morningstar2019,Dumitrescu2018,Goremykina2018,Zhang2016,Thiery2017} and a numerical 
study~\cite{Loic2018}.
Dumitrescu\cite{Dumitrescu2018} {\it et al.} foresee a Kosterlitz-Thouless-type scenario
that also exhibits an intermediate MBL phase; 
also in this case an exponentially diverging length scale was predicted at the transition into the ergodic phase 
~(implying $\nu{=}\infty$), 
however with an exponent $|W-\Wc|^{-1/2}$ (rather than $|W-\Wc|^{-1}$).  
This intermediate phase is characterized by power-law distributed rare thermal 
blocks with an exponent dependent on disorder strength, taking a universal value only at the critical 
point~\cite{Goremykina2018, Dumitrescu2018, Loic2018}; at strong disorder the distribution becomes exponential. 

In most recent work\cite{Morningstar2019} a modification 
of the RG scheme underlying Ref. \onlinecite{Goremykina2018} was proposed,
introducing a second length scale. The intermediate phase 
of the original model appears to be unstable under this modification and, in fact, a power-law 
distribution occurs only at the critical point. 
Therefore, from the perspective of phenomenological RGs,
the existence of an intermediate phase is still under debate.
% 
%% A picture invoking rare thermal blocks, 
%% i.e. a Griffith region within the MBL phase, 
%% has already been proposed in earlier phenomenological 
%% renormalization group settings~\cite{Zhang2016, Thiery2017}. 
In order to clarify if creep relates to rare disorder fluctuations, 
we have been comparing dynamics in random and quasi-periodic potentials in Fig.~2, 
see also the appendix. % supplemental information~\cite{SuppMat}. 
Remarkably, we observe a qualitatively similar slow dynamics in both cases. 
This points towards a possibility that rare events are not the only viable 
explanation for creep as observed, e.g., in Fig.~3.

Our assertion of slow dynamics receives 
(tentative) support from most recent experiments in optical lattices~\cite{RispoliExp18}.  
The Bose-Hubbard model was simulated with quasi-periodic disorder potential; 
a slow particle transport has been seen at finite system sizes~($L=8, 12$) 
above the critical disorder strength. We would like to interpret this observation 
as a possible experimental manifestation of a creep-type dynamics. 
%

%%%%%%%%%%%%%
\subsection{Numerical studies.}
%%%%%%%%%%%%
The exponent function $\beta(t)$, Eq. \eqref{e1}, was considered before 
by Bar Lev, Cohen and Reichman.\cite{BarLevPRL2015}
The fitted exponents together with selected raw data are shown in Fig. 3 
of their work. Even though system sizes and observation times 
are considerably smaller than the ones used in our paper, 
the onset of creep at large disorder 
is clearly visible in their raw data, e.g. at 
$W{\approx}7$. 

One would expect that manifestations of the ``creep phenomenon''
exist in the time-dependency of generic correlation functions and 
we believe that this indeed is the case. 
For instance, 
Doggen et al.~\cite{Doggen2018} have simulated the quench of an
initial state with maximum sublattice imbalance. In Fig. 2 
of their work they observe a 
power-law decay of the imbalance with an exponent $\tilde \beta(W)$. 
Specifically, they report an exponential decrease 
that roughly parametrizes as 
$\tilde \beta(W) \approx \tilde \beta(2)\exp(-(W-2))$ in $2<W<5$; 
at larger $W$ the exponent has become so small that 
it can no longer be resolved within the error bars. 
We interpret this result as broadly consistent with our observations; 
in particular, it implies that $W_\text{c}>5$ for the model considered 
and the localized phase proper, strictly speaking, 
is outside of the computational observation window. 
\footnote{With respect to $W_\text{c}$ a similar conclusion 
can also be drawn from the scaling of the 
Schmidt-gap displayed in Fig. 5 of the same work. 
\cite{Doggen2018}}

Gopalakrishnan et.~al.~\cite{GopalakrishnanPRB15} investigated the low-frequency asymptotics of the conductivity. They reported  
$\sigma(\omega)\sim |\omega|^\alpha$ 
with an exponent $\alpha$ varying continuously
within a region in the MBL-phase. 
At stronger disorder the agreement reflects the fact 
that interaction effects are almost invisible in the numerical data, 
so the non-interacting result $\alpha{\sim}2$ is reached trivially
at large $W$ within the available range of computational parameters. 
At weaker disorder near the border to the thermal phase, 
\textcite{GopalakrishnanPRB15} give a smaller value $\alpha{\sim}1$
corresponding to a logarithmic growth of $\Delta x(t)$. This is consistent
with the behavior we observe in \mblA (see Fig.~2). In fact, a more recent study~\cite{SteinigewegPRB16} on significantly 
larger system sizes reports 
$\alpha \sim 1$ in a broad window of disorder strengths near the transition. 
We mention that an exponent varying continuously 
across the transition, $\alpha(W)$, has also been found in a self-consistent perturbation theory in 
disorder~\cite{PrelovsekPRB17}.
% {\tt I forgot: Why do we say "however"?} 
% 

Similarly, also Bari\v{s}i\'c et. al.~\cite{PrelovsekPRB16} 
report values $\alpha{\sim}1$ in the MBL-phase, employing 
system sizes up to $L{=}28$. 
In their fitting, the authors are allowing for a non-vanishing 
residual dc-conductivity, however, and indicate that finite 
size corrections are not fully under control. So, we understand 
that the numerical value $\alpha{=}1$ is subject to a 
significant uncertainty.
The same group later extended their studies to related 
correlation functions~\cite{MierzejewskiPRB16, Prelovsek-review-2017}. 
Reminiscent of creep, also in these correlators a 
very slow dynamics with significant finite-size effects can be identified, 
e.g., in Fig.~2 (c,d) of \textcite{MierzejewskiPRB16}. 

We mention that the sparsity of many-body states in Fock-space can give rise to 
a kind of multifractality, which exhibits a non-universal and, in general, 
basis dependent spectrum~\cite{Bogomolny2014, Gornyi2017}. 
Most recent numerical work has confirmed this earlier result at least 
within a certain window of system sizes computationally available~\cite{MaceMultifractality2018}. 
If and how multifractality of the many-body wavefunction manifests 
in the evolution of $\alpha(W)$ remains to be clarified. 

Finally, also the study by Serbyn, Papic and Abanin exhibits 
discernible finite size effects, which may be consistent with creep~\cite{Serbyn2015}. 
Specifically, these authors study expectation 
values of energy-normalized overlap matrix elements of local operators. 
Fig. 3 of this work shows that with increasing system size, $L$, the flow of the 
average turns around from insultating and redirects towards thermalizing 
behavior. The system size of turn-around, $L^*(W)$, is seen to grow with disorder. 
Since in this work only system sizes up to $L{=}16$ have been studied, 
it remains open if a finite critical value of disorder, $W_\text{c}$, exists 
at which $L^*(W)$ diverges, and what numerical value it takes. 
\footnote{We express our gratitude to Ehud Altman for drawing our attention 
to this work.}

%%%%%%%%%%%%%%%%%%%%%%%%%%%%%%%%%%%%%%
\begin{figure}[t]
 % \centering  
  \includegraphics[width=1.0\columnwidth]{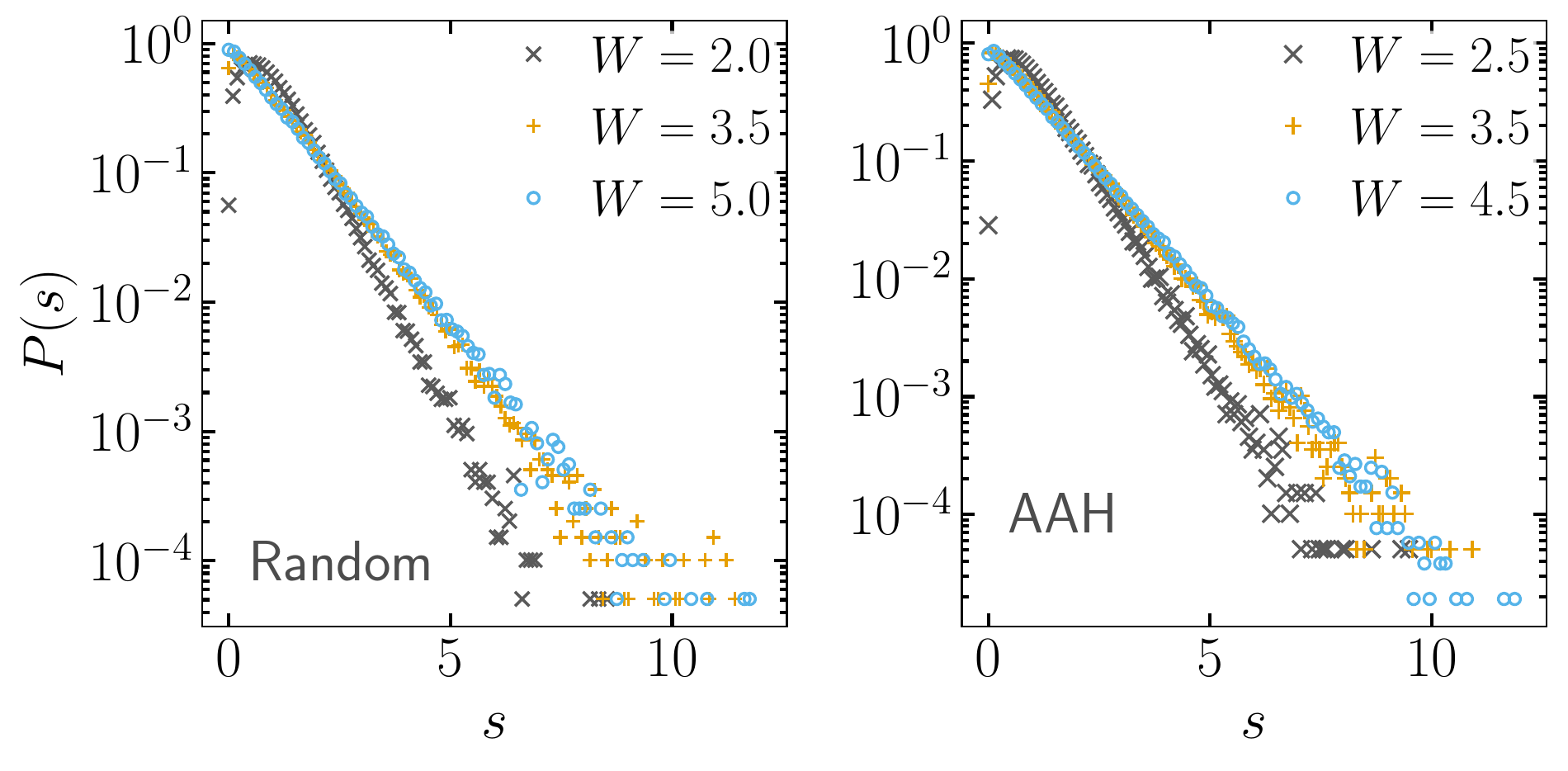}
  \caption{Level spacing distribution for the $t-V$-model at interaction strength $V{=}2\thop$ for the fully disordered 
model (left, $W/\thop{=}{2.0, 3.5, 5.0}$) and quasi-periodic (AAH) model (right, $W/\thop{=}{2.5, 3.5, 4.5}$) at 
system size $L{=}16$. Both panels show an evolution of the distribution from Wigner distribution towards Poisson 
statistics with increasing disorder. 
  \label{Sf0}  }
\end{figure}
%%%%%%%%%%%%%%%%%%%%%%%%%%%%%%%%%%%%%%
%%%%%%%%%%%%%%%%%%%%%%%%%%%%%%%%%%%%%%%%%%%%%%%%%%%%%%%%%%%%
\subsection{Further discussion: Spectral statistics and creep \label{si1}} 
%%%%%%%%%%%%%%%%%%%%%%%%%%%%%%%%%%%%%%%%%%%%%%%%%%%%%%%%%%%%
\subparagraph*{Level spacing distribution $P(s)$.} 
We comment on the spectral 
statistics of the random potential and the AAH model. 
Figure~\ref{Sf0} shows the evolution of the level spacing distribution 
$P(s)$ from weak~(Wigner-Dyson statistics) 
to strong~(Poisson distribution) disorder as it is observed in a 
relatively small system of size $L{=}16$. 
The data is qualitatively similar to the one presented, 
e.g., in Ref.~\onlinecite{Bertrand2016}.
It suggests that at disorder values exceeding 
$W\approx 3$ the $P(s)$ distribution is nearly Poisson and 
therefore samples are in a certain spectral sense 
close to the many-body localized phase. 
As is also seen from the Fig.~\ref{Sf0} that statement holds 
irrespective of whether the potential is fully random or quasi-periodic. 

Importantly, the distribution $P(s)$ undergoes an evolution 
with increasing system size $L$, see e.g. 
Fig. 7  in Ref.~\onlinecite{Bertrand2016} and Fig. 9 in Ref.~\onlinecite{SierantPRB19}.  
%{\tt P. Sierant, J. Zakrzewski, Phys. Rev. B 99, 104205 (2019)}. 
At stronger disorder values and with $L$ increasing from small values, e.g. 
$L{\sim}12$, the initial flow is towards a Poisson shape. 
This observation is fully consistent with - and in fact an 
indispensable requirement for - the 
applicability of the concept of local integrals of motion that underlies 
the current understanding of many-body localization.

Notice, however, that at present there is no proof that 
indeed a critical value of the disorder, $W_\text{c}$, exists, 
such that the asymptotic flow of $P(s)$ 
for the $t{-}V$-model or Heisenberg model 
is indeed towards Poisson at disorder values exceeding 
$W_\text{c}$.  
The observation of ``creep''  indicates that the critical 
disorder strength, $W_\text{c}$, may be very large; if it is finite at all, 
then it must correspond to a non-interacting localisation length that is 
considerably smaller than the lattice spacing. 

A conclusive numerical study of $P(s)$ to this end is lacking as of today. However, 
inspecting the data for $P(s)$ shown in Fig. 9 of Ref.~\onlinecite{SierantPRB19}
at very small $s$ and  disorder, $W{=}4$,  
the flow towards Poisson indeed appears to stop 
already near $L\approx 18$. Conceivably, at larger $W$ 
a similar termination can be observed, 
if only at larger system sizes $L(W)$.
Therefore, the present evidence concerning $P(s)$ being 
close to Poisson at smaller system sizes is not inconsistent with the creep-scenario. 

%\textcolor{red}{\tt We need to define the AA Hamiltonian to check factors of two.} 
% The data is shown for both the random and AAH model for $L=16$ and $V/\thop=2$. 

\subparagraph*{Level statistics ratio $r_n$.} 
A simplified diagnostics of the level statistics is provided  by the 
ratio $r_n(W,L){\coloneqq}\text{min}(\Delta_n,\Delta_{n+1})/\text{max}(\Delta_n,\Delta_{n+1})$,
where $\Delta_n{\coloneqq} E_n - E_{n+1}$ 
is the spacing between eigenvalues.
For the Gaussian-orthogonal ensemble 
(nearly chaotic situation, $W{\ll} W_\text{c}$) 
one has $r_n{\simeq}0.53$ while for the Poisson case 
(localized situation,  $W{\ll} W_\text{c}$)  
$r_n{\simeq}0.39$. In analogy to conventional second-order 
phase transitions, it is tempting to test for  single-parameter
scaling near the critical  point, i.e. $r_n(\xi/L)$, defining the localization length, 
$\xi\propto |W-W_\text{c}|^{-\nu}$, with the corresponding exponent 
$\nu$. Indeed, the earlier attempts to determine $W_\text{c}$ and 
$\nu$ did employ such an analysis~\cite{Luitz2015}. 

%\SB{In lower panel of Fig.~\ref{Sf0} we show such a scaling collapse for $V/\thop=2.0$ for system sizes %$L=\{12,14,16,18\}$ for the random model. This clearly shows that for these system sizes a reasonably %good collapse is achievable with the following parameters $\nu \approx 1.1$ and $W_c \approx 3.8$. This %way of analyzing the data underestimate the $W_c$ as also pointed out in some other works.}

It has been pointed out by Khemani {\it et al.}~\cite{Khemani2017} already that
due to the small system sizes available, 
-- typically not exceeding $L{=}20$, -- 
one is stuck in a pre-asymptotic regime. A clear illustration 
indicating how large the actual distance to criticality really is 
can be drawn, e.g., from Fig. 2b of this Ref.~\onlinecite{Khemani2017}. 
A pair of traces of $r_n(W,L)$ for
$L{=}12,14$ intersects near $W{\approx} 5.5$ while the pair  
$L{=}16,18$ intersects near $W{\approx}7$. The true scaling limit 
would announce itself by all pairs intersecting at the same point 
$W{=}W_\text{c}$.  

Attempting in this situation a scaling analysis and fully ignoring finite size 
corrections yields $\nu{\approx}1$. The result does not violate the Harris criterion,  
because the validity of the latter is restricted to the critical point, 
while the scaling analysis has been done in a preasymptotic regime. 
Following the same type of analysis, one expects with increasing system sizes 
a drift of $\nu$ and $W_c$ to ever larger values. 
Again, the existing data seems consistent with the creep scenario. 

We mention that a drift in pair-wise intersection 
points is very common in computational studies of Anderson transitions. 
For a recent example how to include finite-size corrections in 
a case relevant to MBL, i.e. random-regular graphs, 
we refer to a recent study by Ref.~\onlinecite{TikhonovRRG16}.

\section{Conclusion} 
In this work we have studied the density autocorrelation function of 
spinless fermions in the $t-V$-model at infinite temperature. 
We detect strong finite-size effects and 
a very slow dynamics  - ``creep'' - with a tendency towards 
thermalization even at strong disorder, where the level statistics is 
close to Poissonian. Our observations are very similar for truly random 
and quasi-periodic potentials, so that rare-region effects seem discouraged 
as a likely origin of subdiffusion and creep. We believe that our results  
imply one out of two possibilities: 

(i)  ``Creep'' is a transient phenomenon, perhaps indicative of 
the critical fan~\cite{Khemani2017} 
(similarly for quasi-periodic potential~\cite{Setiawan2017}). 
In this case, the width of a wavepacket, $\Delta x(t)$,
either grows in a power law manner, $\Delta x(t){\sim}t^{\beta}$,
in a thermalizing phase, or 
exhibits a logarithmic (or weaker) growth window
in time after which $\Delta x(t)$
eventually saturates at a finite value, $\xi(W)$, in the MBL-phase.

If this is the case, an important conclusion of our study is that due to creep  
the asymptotic regime, $\Delta x(t)\approx \xi(W)$, in simulations is 
very difficult to reach in a controlled manner. We find that even at disorder values 
such that the non-interacting localization length is considerably smaller 
than the lattice spacing, $\Delta x(t)$ does not saturate within the 
simulation window of $1000 \thop^{-1}$. 
Hence, a numerical study of the true asymptotics including $\xi(W)$ 
- especially with respect to criticality - 
may have been out of reach of simulations of physical observables thus far, 
and, presumably, even in experiments~\cite{AbaninBloch-Review-2018}.

(ii) An alternative possibility is that ``creep''
may indicate the existence of marginally thermalizing phases, 
such as the subphase \mblA discussed in this work. The characterizing feature is that 
$\Delta x(t)$ keeps growing in time weaker than any power, 
 eventually crossing any finite upper bound $\xi$.  
In this case, the numerical observation of creep-type phenomena 
would be closer to the asymptotic regime, 
but the present understanding of the MBL-phase would be incomplete. 
Conceivably, such phases can be described
by complementing the concept of local integrals of motion\cite{Serbyn2013,Huse2014}, 
which has been developed for the case $W{>}W_\text{c}$, 
with a weak integrability breaking mechanism.

\begin{acknowledgments}
{\bf Acknowledgments.}~We thank  E. Altman, I. Gornyi, V. Khemani, Z. Lenarcic, J. Moore, A. Mirlin and D. Polyakov for inspiring discussion. FE  expresses his sincere gratitude to Ehud Altman and the team at UC Berkeley 
for their generous hospitality . 
SB would like to thank G. De Tomasi for earlier collaboration on a similar topic and for many insightful discussions. SB acknowledges support from DST, India,
through Ramanujan Fellowship Grant No. SB/S2/RJN-128/2016 and ECR/2018/000876; FE was supported from
the DFG under Grants No. EV30/11-1 and EV30/12-1. We acknowledge computing time on the supercomputer
ForHLR funded by the Ministry of Science, Research and the Arts Baden-W\"urttemberg and by the Federal Ministry of Education and Research. 
% {\color{red} Felix - LRZ??, do we} 
\end{acknowledgments}

%%%%%%%%%%%%%%%%%%%%%%%%%%%%%%%%%%%%%%%%%%%%%%%%%%%%%%%%%%%%
\appendix

%%%%%%%%%%%%%%%%%%%%%%%%%%%%%%%%%%%%%%%%%%%%%%%%%%%%%
\section{Time dependence of $\Phi(x,t)$}
\begin{figure}[htb]
 \centering  
  \includegraphics[width=1.0\columnwidth]{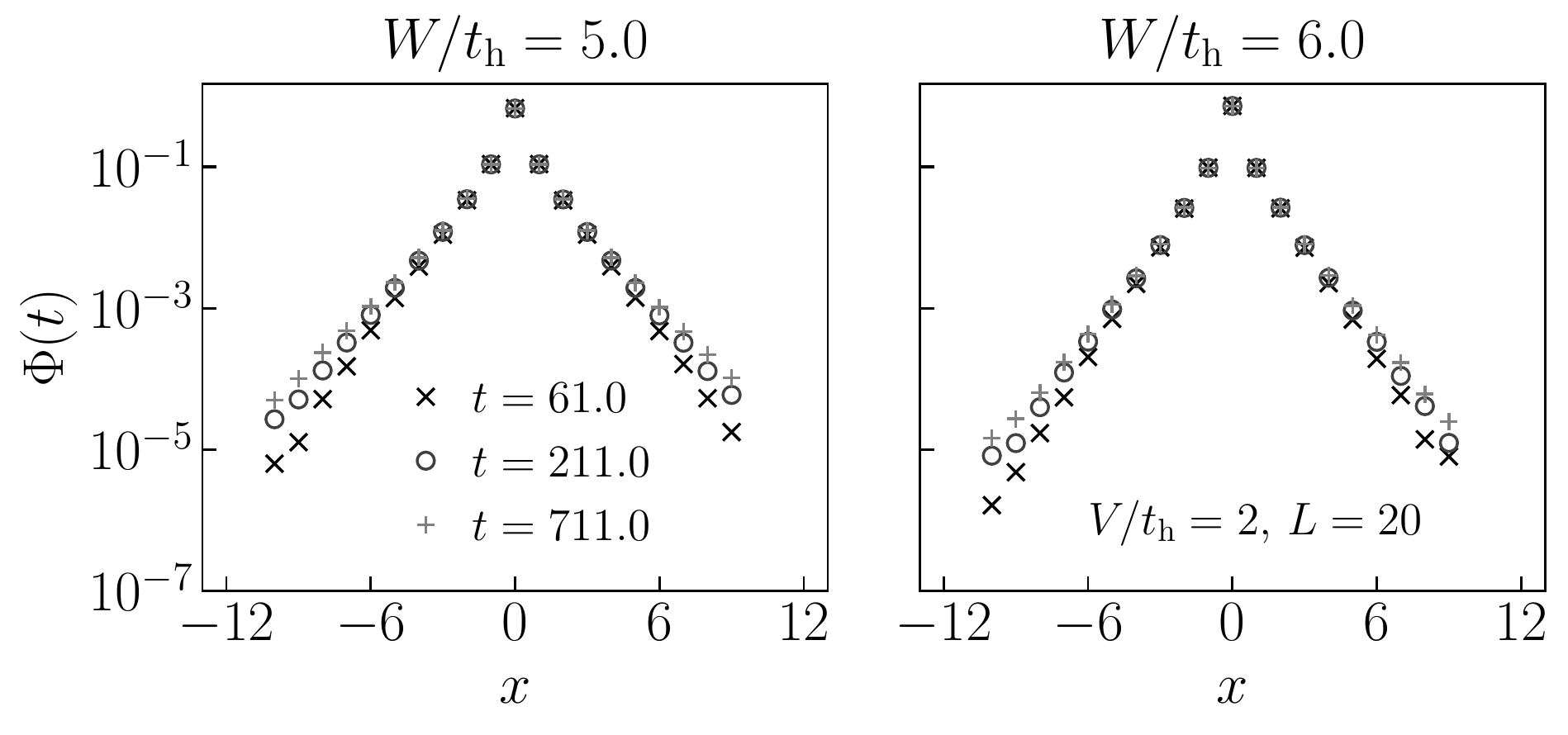}
  \caption{Time evolution of the full distribution function $\Phi(t)$ for the random model at strong disorder. 
  The slow dynamics manifests as tails lifting away from the nearly exponential shape that is indicative 
  of the short time behavior and the non-interacting limit. 
  The data is shown for $L=20$, $V/\thop=2$ and two different diorder values $W/\thop=5.0, 6.0$.  
  \label{Sf2:dist}  }
\end{figure}

%%%%%%%%%%%%%%%%%%%%%%%%%%%%%%%%%%%%%%%%%%%%%%%%%%%%%
In the localized phase $\dex (t)$ increases slowly~(as shown in Fig.~2) with time above the non-interacting saturation  
within our observation time~($\sim 10^3$ in unit of inverse $\thop$). In the distribution function 
$\Phi(x,t)$ this propagation is reflected via the lifting of the tails with time as seen in Fig.~\ref{Sf2:dist}. 
The tail-lifting eventually produces a stretched-exponential shape of $\Phi(x,t)$. 
% \textcolor{red}{This slow propagation in the localized phase is also observed in the AAH 
% model albeit it seems to saturate faster with increasing disorder. 
% {\tt What does this follow from??}} 
%%%%%%%%%%%%%%%%%%%%%%%%%%%%%%%%%%%%%%%%%%%%%%%%%%%%%

%%%%%%%%%%%%%%%%%%%%%%%%%%%%%%%%%%%%%%%%%%%%%%%%%%
\section{Oscillations in $\beta$ \label{si2} \label{sec:oscillations-beta}}
%
%%%%%%%%%%%%%%%%%%%%%%%%%%%%%%%%%%%%%%%%%%%%%%%%%%%%%%%%%%%%
\begin{figure}[b]
 % \centering  
  \includegraphics[scale=0.30]{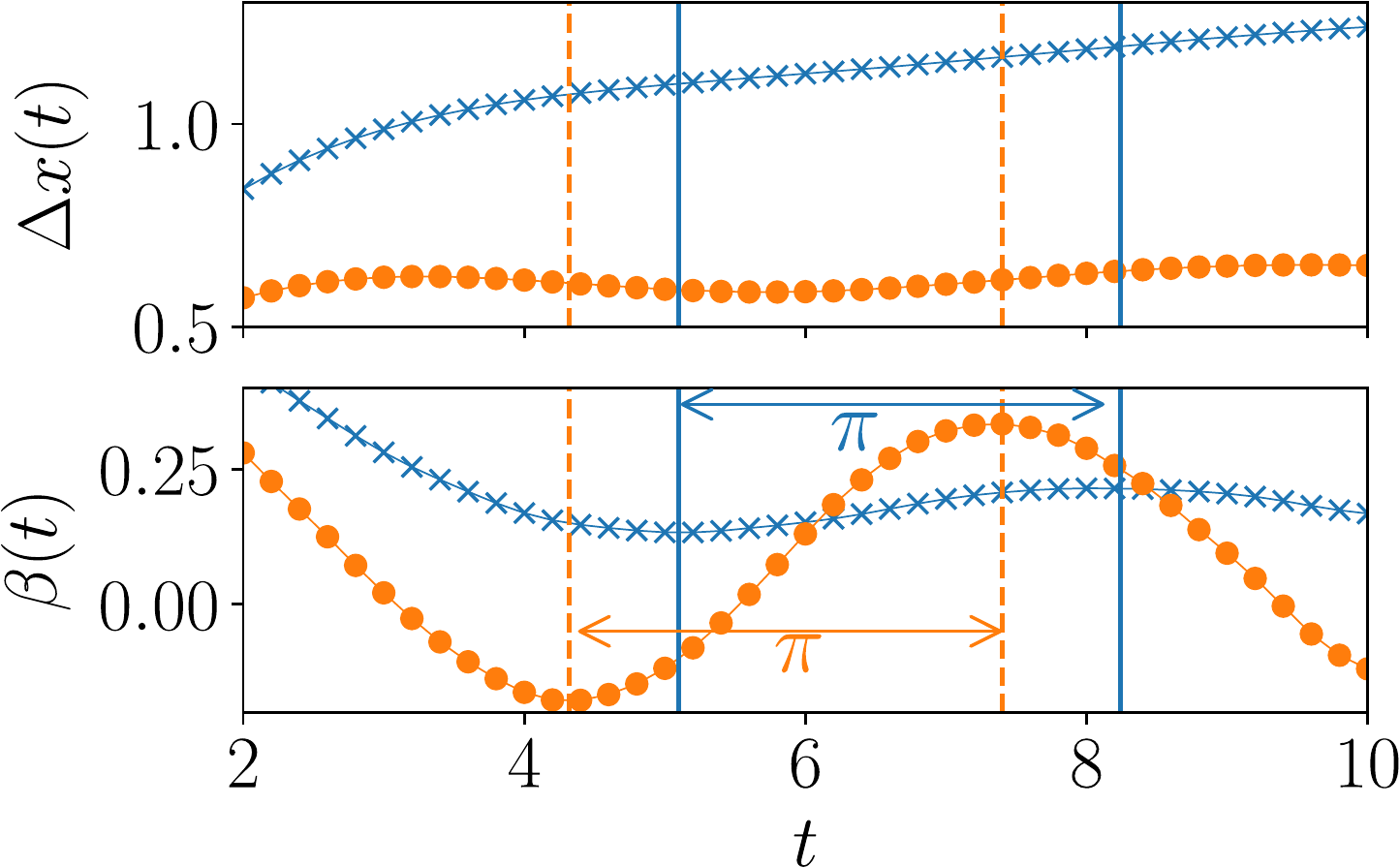}
  \includegraphics[scale=0.29]{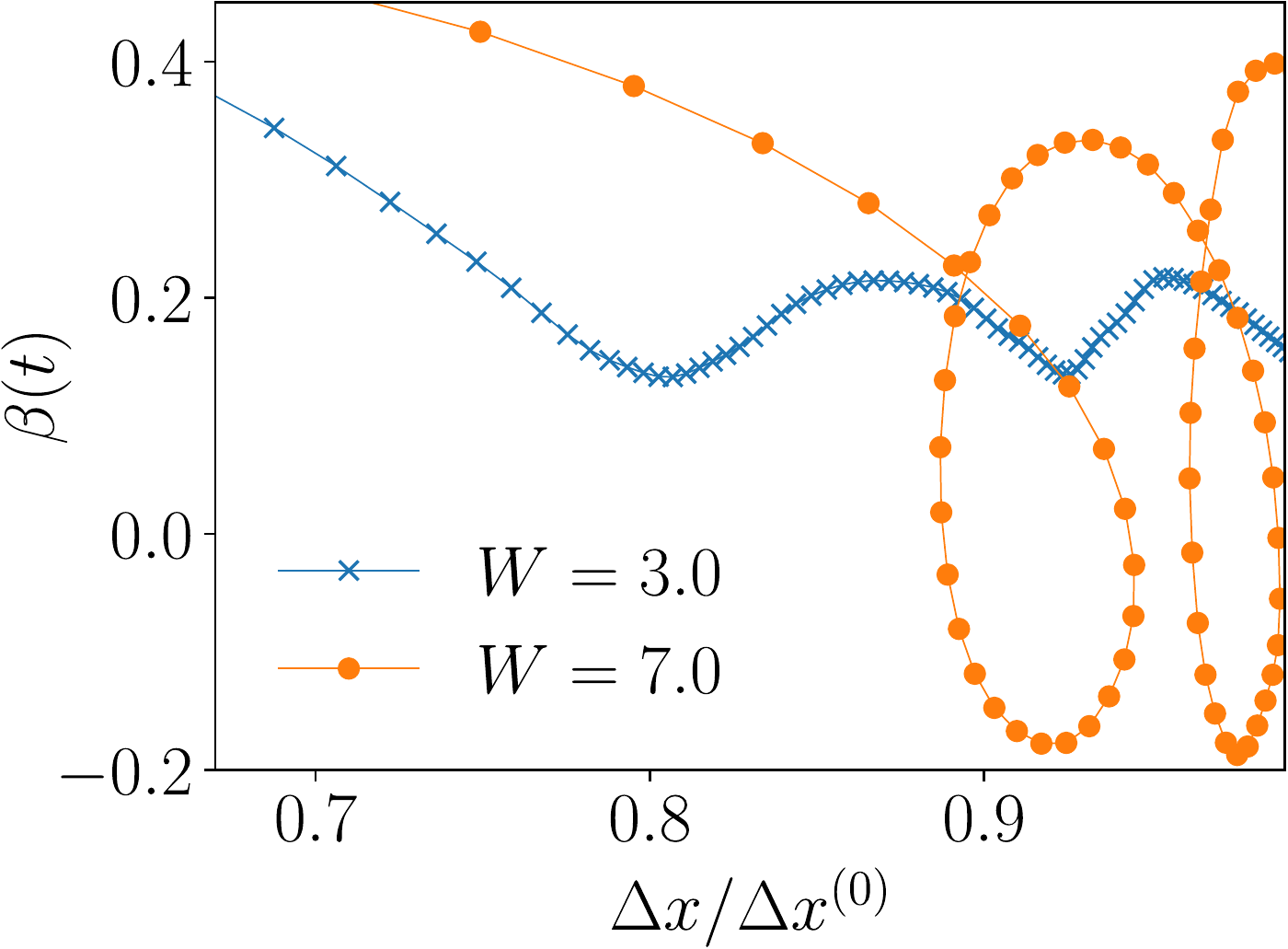}
  \caption{Short-time behavior of $\Delta x(t)$ and $\beta(t)$ averaged over several thousand 
  samples. Figure is  illustrating the origin of the peak-structure seen in 
  Fig.~3 of the main text
  at intermediate (blue) and strong (orange) disorder: $W{=}3.0,7.0$. 
  Left: $\Delta x(t)$ exhibits oscillatory behaviour that is emphasised in its logarithmic derivative, $\beta(t)$. 
  At strong enough disorder, $W{>}W^*$, $\Delta x(t)$ becomes non-monotonous.   
    Vertical lines indicate a half-period of the oscillation. 
  Right: Multivalued  $\beta(\Delta x)$ at $W{>}W^*$.
  \label{Sf1}  }
% \twocolumngrid
\end{figure}
%%%%%%%%%%%%%%%%%%%%%%%%%%%%%%%%%%%%%%%%%%%%%%%%%%%%%%%%%%%%
We analyze the spike observed in $\beta(\dex)$ (see Fig.~3), 
which becomes more pronounced at increasing disorder strenth $W$. 
It occurs at $\Delta x \lesssim 1$ (cf. Fig.~\ref{Sf1}), and therefore 
we interpret it as a signature of the lattice spacing, which is is taken as the unit of length. 
In essence, the spike is a manifestation of strong oscillations that induce 
a non-mononous growth of $\Delta x(t)$ in time. 
Such oscillatory behaviour arises from neighboring pairs of sites 
that have a common local potential, which deviates strongly from the environment 
and thus is forming a "trap".
%%%%%%%%%%%%%%%%%%%%%%%%%%%%%%%%%%%%%%%%%%%%%%%%%%%%%
\begin{figure}[t]
 % \centering
  \includegraphics[scale=0.6]{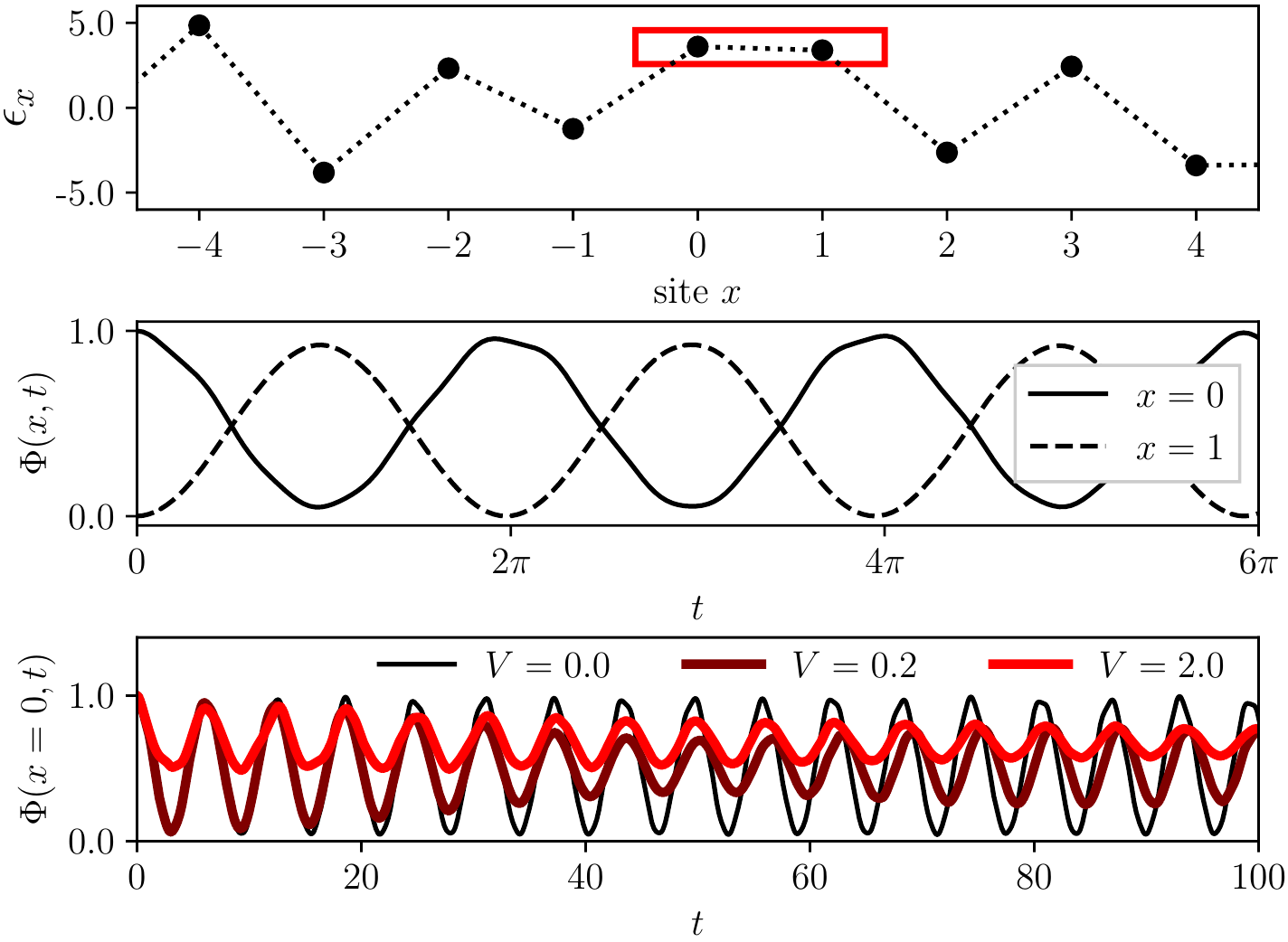}
  \caption{Behavior of $\Phi(x,t)$ around $x=0$ for a given disorder configuration. 
  Note that $\Phi(x,t=0)=\delta_{x,0}$, by definition. 
  Top: Disorder potential $\epsilon_x$ around $x=0$: sites $x=0,1$ are very close in energy.  
  Center: Coherent oscillations between neighboring sites $x=0,1$ in the non-interacting limit $V=0$. The period is 
close to $2\pi$. 
  Bottom: Damping for finite interaction strength $V$. 
  \label{Sf2}
  }
\end{figure}
%%%%%%%%%%%%%%%%%%%%%%%%%%%%%%%%%%%%%%%%%%%%%%%%%%%%%
%
The trap constitutes a two-site system with an associate energy doublet that is split by the hopping term. 
The dynamics of the associated charge density exhibits a characteristic frequency $\omega \approx  \thop = 1$. The corresponding oscillations
in the correlator $\Phi(x,t)$ are illustrated in Fig.~\ref{Sf1} 
for a given disorder configuration with trap near $x{=}0$.  
Strong oscillations are also seen in Fig.~\ref{f0} differing 
from Fig.~\ref{Sf2} by rescaling the 
on-site potential to smaller values with factor $3/5$ (blue) and $1.5/5$ (red).

We discuss the effect of interactions on time-dependent fluctuations in $\Delta x(t)$: 
In the non-interacting limit, $V{=}0$, pronounced oscillations are seen  large disorder (corresponding to strength 
$W{=}2$) 
in Fig. \ref{Sf2}; they persist up to very long times. With decreasing disorder, Fig.~\ref{f0}, 
lower frequencies are typically mixing in because resonances between more distant sites become 
accessible, converting oscillations to mesoscopic fluctuations. 
This type of oscillation survives the ensemble average, because 
return times tend to be integer multiples of $2\pi/\thop$. 
% {\color{red} FW, please check $\thop$}. 

With finite interaction the random potential will, in general, undergo a static 
renormalization; it shifts fluctuation frequencies without affecting the qualitative physics. 
New  effects can be expected from dynamical renormalizations, i.e., from dephasing; 
it is a characteristic feature of many-body systems expressing the possibility of particles 
to scatter off each other and thus exchanging momentum and energy. 
In the correlator $\Phi(x,t)$ dephasing manifests itself in a damping of the mesoscopic fluctuations, 
see Fig.~\ref{Sf2}, bottom and \ref{f0}.  
The defining characteristics of the mbl-phenomenon is that despite mesoscopic fluctuations 
washing out very quickly with increasing $V$, 
the average return probability is seen to keep values of order unity that hardly decay at all, 
see Fig. \ref{Sf2} and \ref{f0}.

%%%%%%%%%%%%%%%%%%%%%%%%%%%%%%%%%%%%%%%%%%%%%%%%%%
\section{Evolution of $\Delta x(t)$ in a quasi-periodic  (AAH) potential \label{si3}} 
%%%%%%%%%%%%%%%%%%%%%%%%%%%%%%%%%%%%%%%%%%%%%%%%%%
%%%%%%%%%%%%%%%%%%%%%%%%%%%%%%%%%%%%%%%%%%%%%%%%%%%%%
\begin{figure}[t]
%  \centering
  \includegraphics[scale=0.45]{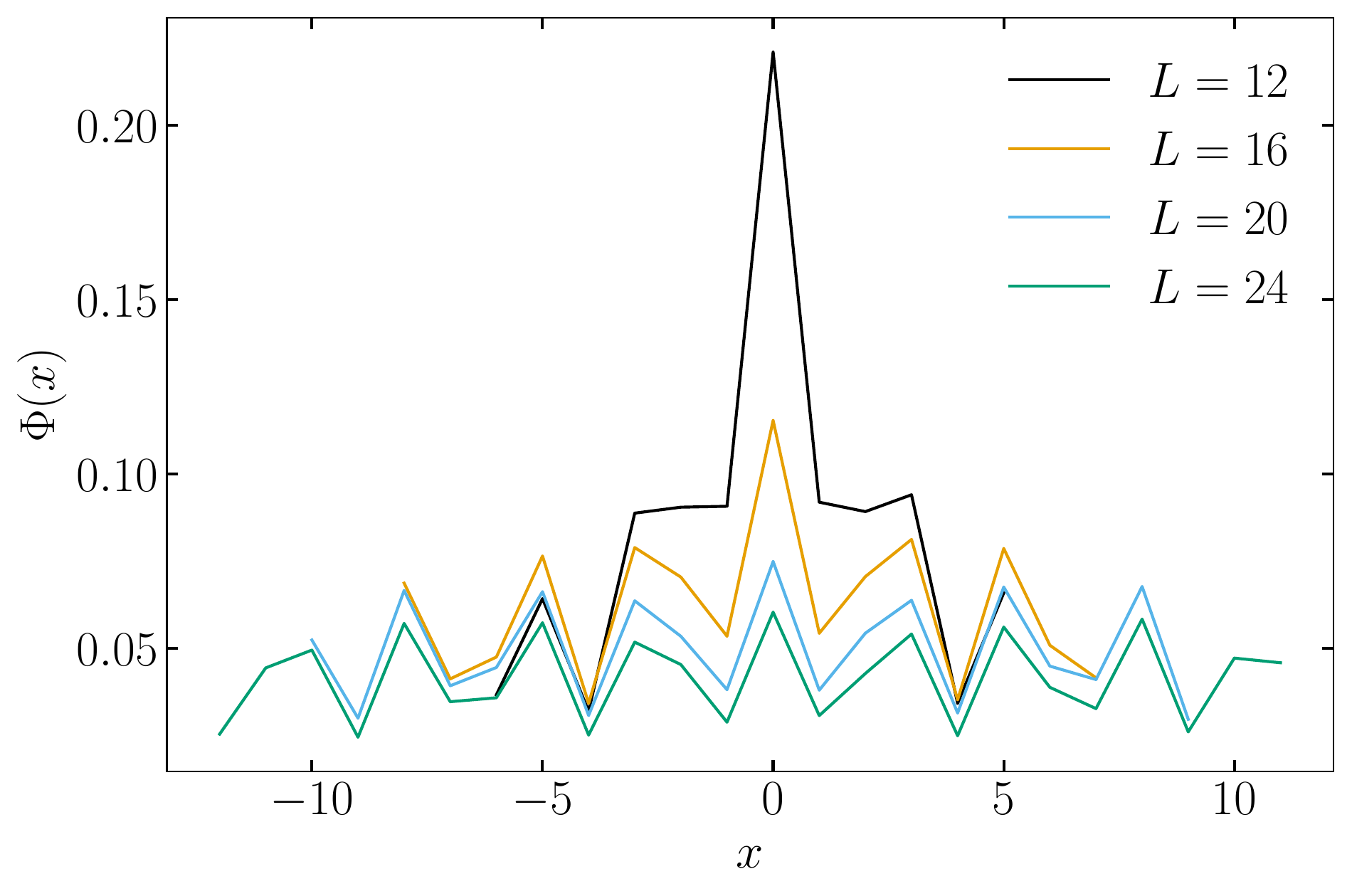}
  \caption{$\Phi(x,t)$ at moderate strength of the quasi-periodic potential 
  displayed at a very large observation time $t\sim 10^3\thop^{-1}$ 
  for system sizes $L{=}12,16,20,24$. 
%   {\tt Please, cut away the top text} 
  \label{Sf3b}} 
\end{figure}
%%%%%%%%%%%%%%%%%%%%%%%%%%%%%%%%%%%%%%%%%%%%%%%%%%%%%
In the main text we have analysed the evolution of the correlator $\Phi(x,t)$ for the case of 
fully disordered potentials, Fig.~3. We here display the analogous data  
obtained for the quasi-periodic case. 

{\bf Thermalizing regime.} 
% 
%%%%%%%%%%%%%%%%%%%%%%%%%%%%%%%%%%%%%%%%%%%%%%%%%%%%%
\begin{figure*}[t]
\centering
\includegraphics[width=1.0\textwidth]{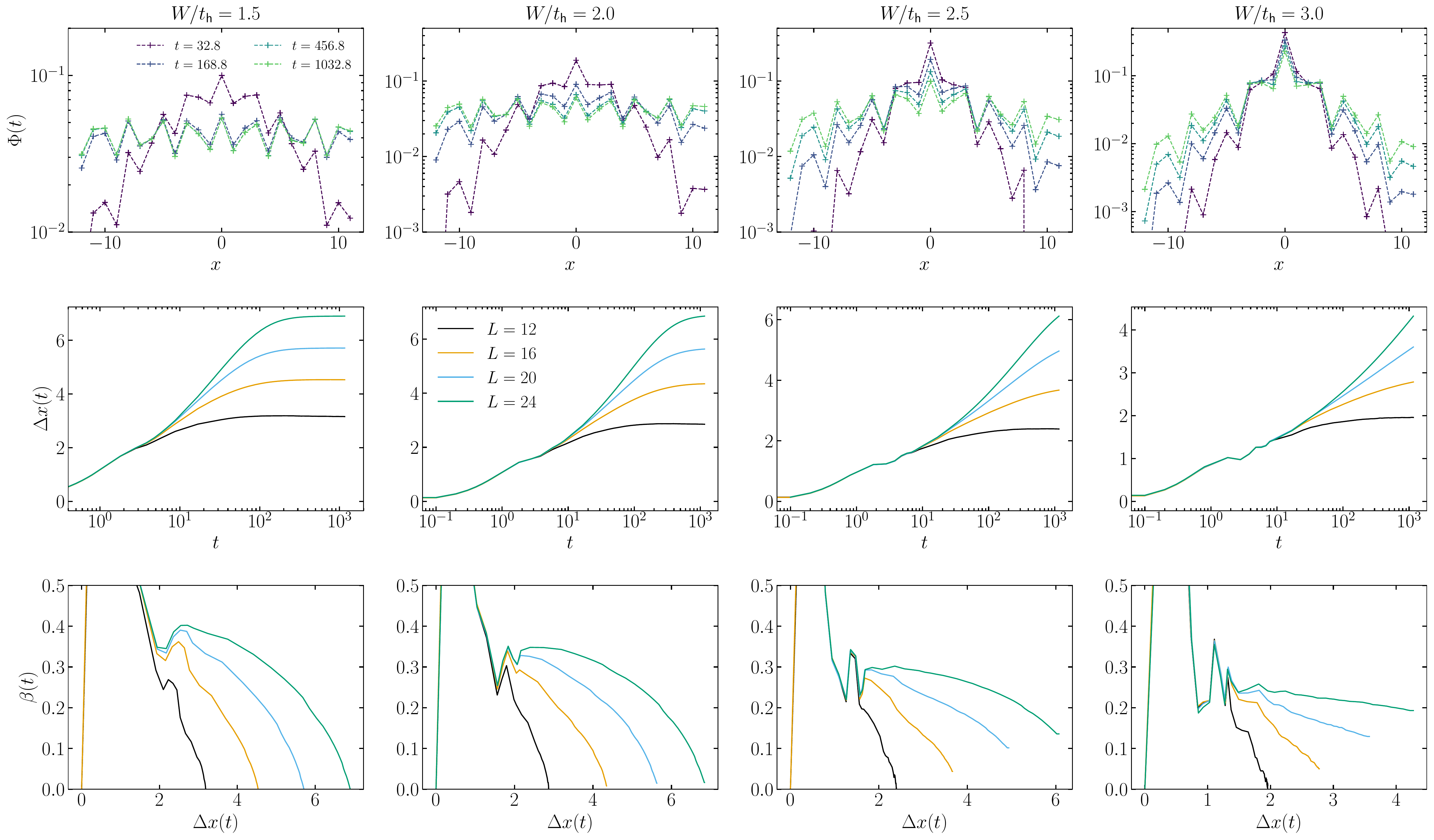}
\caption{Upper panel: Shows the evolution of $\Phi(t)$ for different values of disorder~($W/\thop=1.5, 2.0, 2.5, 3.0$) 
at interaction $V/\thop=2.0$ for the AAH model for $L=24$. Inset (b) displays the long time density correlator that 
shows pronounce oscillation, which is reminiscent of the underlying quasi-peroidic potential~(black dashed line).  
Middle row shows the corresponding $\Delta x(t)$  and the lower panel displays the evolution of $\beta(t)$ with system 
sizes~($L=12, 16, 20, 24$) for different disorder values. 
  \label{Sf8}} 
\end{figure*}
%%%%%%%%%%%%%%%%%%%%%%%%%%%%%%%%%%%%%%%%%%%%%%%%%%%%%
Before we discuss the main data set, Fig.~\ref{Sf8}, we report an effect 
that we observe for quasi-periodic potentials, which is not seen in the fully random case. 
In Fig.~\ref{Sf3b} we show the correlation function $\Phi(x,t)$ at very large times. 
Despite the fact that the dynamics has explored the full system size, 
oscillations are seen that do not show a tendency to equilibrate, 
see also Fig.~\ref{Sf8} for other disorder values. 
While a detailed study of this effect is beyond the scope of this 
work, a few comments are in place. 

A preliminary analysis suggests that the 
amplitude of the oscillations decays with increasing system size in a power
law-fashion; therefore, we tentatively consider them  as a finite
size effect. Presumably, the oscillations are related to the correlations in the 
underlying quasi-periodic potential; hence, one would expect that their 
shape is sensitive to the choice of the parameter $\alpha$ in the definition 
of $\epsilon_x$~(see~\footnote{When working with quasi-periodic potential 
(Aubry-Andr\'e model)  we choose, 
$\epsilon_x {=} W \cos(2\pi \alpha x + \phi)$ with $\alpha=2/(\sqrt{5}+1)$
and $\phi$ is the random phase distributed uniformly between $[0, 2 \pi]$} for further details). 
% \textcolor{red}{Here I would like to refer to the footnote.. } 
%
Furthermore  our data Fig.~\ref{Sf8} indicates that due to these oscillations 
at moderate disorder $W{=}1.5-3.0$,
i.e. in a regime that is believed to be thermal, 
finite size effects can be strong.  
We believe that they should be observable, in principle, 
also in typical cold-atom experiments~\cite{Luschen2017}.
Finally, in recent computational work it was found that the critical 
disorder strength could also be dependent on $\alpha$: 
$W_\text{c}(\alpha)$~\cite{Doggen2019}.
%\textcolor{red}{Doggen, Mirlin 2019} 
Conceivably, this finding is related to effects such as displayed 
in Fig.~\ref{Sf3b}. 
 %%%%%%%%%%%%%%%%%%%%%%%%%%%%%%%%%%%%%%%%%%%%%%%%%%%%%
\begin{figure*}[bht]
 \centering  
  \includegraphics[width=1.0\textwidth]{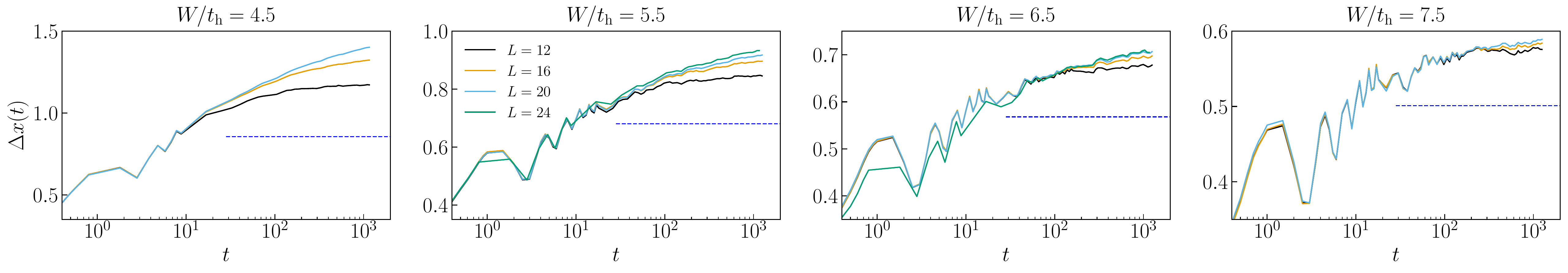}
  \caption{Shows the time evolution of $\dex(t)$ for different values of disorder strength $W/\thop=4.5, 5.5, 6.5, 7.5$ 
and for interaction strength $V/\thop=2.0$. The dashed line (blue) is indicating the 
  non-interacting value of $\dex(t)$ for corresponding disorder strengths for $L=16$ obtained from separate 
computations. \label{Sf:locAAH}  }
\end{figure*}
%%%%%%%%%%%%%%%%%%%%%%%%%%%%%%%%%%%%%%%%%%%%%%%%%%%%%

The analysis of $\Phi(x,t)$ for the quasi-periodic case in terms of the effective exponent $\beta(t)$ is done in analogy 
to the fully random case. 
The middle panel of Fig.~\ref{Sf8} displays the evolution of $\dex(t)$ at system sizes 
$L{=}12,16,20,24$ for interaction $V/\thop=2.0$.  At small disorder, $\dex(t)$ is seen to reach a plateau value $\approx 
L/3.5$ indicating saturation due to the finite system size.
While the saturation width is only slowly decreasing with disorder, 
$W$, the saturation time is clearly seen to be a strongly growing function of $W$ exceeding the computational time 
window  already at moderate disorder $W{=}2.5$. 

The lower panel shows the corresponding evolution of the logarithmic derivative of $\dex(t)$, i.e. the exponent function 
$\beta(t)$. As announced before, the evolution for the quasi-periodic case 
is analogous to the random disorder case~(see Fig.~3 lower panel). 
In particular, extremely strong finite size effects hamper a reliable extraction of the asymptotic exponent. 
The overall trend is that the short-time exponent at weak disorder takes larger values 
which are not far from the diffusive regime, $\beta{=}1/2$. 

{\bf Strong disorder.} 
At stronger disorder the level statistics is seen to approach the Poissionian limit in~(Fig.~\ref{Sf2}). 
We thus associate with disorder values $W\gtrsim 5$ a regime close or in the many-body localized phase. 
Correspondingly, with increasing disorder the dynamics  as displayed by $\dex(t)$ 
slows down considerably, see Fig.~\ref{Sf:locAAH}; in the quasi-periodic and the fully random situations 
it shows the same qualitative behavior. Since effects of rare regions are not expected in the 
quasi-periodic situation, our observations suggest that the typical strong disorder effects may not be the origin of 
slow dynamics and creep. 

\bibliography{MBL}

\begin{thebibliography}{74}%
\makeatletter
\providecommand \@ifxundefined [1]{%
 \@ifx{#1\undefined}
}%
\providecommand \@ifnum [1]{%
 \ifnum #1\expandafter \@firstoftwo
 \else \expandafter \@secondoftwo
 \fi
}%
\providecommand \@ifx [1]{%
 \ifx #1\expandafter \@firstoftwo
 \else \expandafter \@secondoftwo
 \fi
}%
\providecommand \natexlab [1]{#1}%
\providecommand \enquote  [1]{``#1''}%
\providecommand \bibnamefont  [1]{#1}%
\providecommand \bibfnamefont [1]{#1}%
\providecommand \citenamefont [1]{#1}%
\providecommand \href@noop [0]{\@secondoftwo}%
\providecommand \href [0]{\begingroup \@sanitize@url \@href}%
\providecommand \@href[1]{\@@startlink{#1}\@@href}%
\providecommand \@@href[1]{\endgroup#1\@@endlink}%
\providecommand \@sanitize@url [0]{\catcode `\\12\catcode `\$12\catcode
  `\&12\catcode `\#12\catcode `\^12\catcode `\_12\catcode `\%12\relax}%
\providecommand \@@startlink[1]{}%
\providecommand \@@endlink[0]{}%
\providecommand \url  [0]{\begingroup\@sanitize@url \@url }%
\providecommand \@url [1]{\endgroup\@href {#1}{\urlprefix }}%
\providecommand \urlprefix  [0]{URL }%
\providecommand \Eprint [0]{\href }%
\providecommand \doibase [0]{http://dx.doi.org/}%
\providecommand \selectlanguage [0]{\@gobble}%
\providecommand \bibinfo  [0]{\@secondoftwo}%
\providecommand \bibfield  [0]{\@secondoftwo}%
\providecommand \translation [1]{[#1]}%
\providecommand \BibitemOpen [0]{}%
\providecommand \bibitemStop [0]{}%
\providecommand \bibitemNoStop [0]{.\EOS\space}%
\providecommand \EOS [0]{\spacefactor3000\relax}%
\providecommand \BibitemShut  [1]{\csname bibitem#1\endcsname}%
\let\auto@bib@innerbib\@empty
%</preamble>
\bibitem [{\citenamefont {Basko}\ \emph {et~al.}(2006)\citenamefont {Basko},
  \citenamefont {Aleiner},\ and\ \citenamefont {Altshuler}}]{Basko2006}%
  \BibitemOpen
  \bibfield  {author} {\bibinfo {author} {\bibfnamefont {D.~M.}\ \bibnamefont
  {Basko}}, \bibinfo {author} {\bibfnamefont {I.~L.}\ \bibnamefont {Aleiner}},
  \ and\ \bibinfo {author} {\bibfnamefont {B.~L.}\ \bibnamefont {Altshuler}},\
  }\href {http://www.sciencedirect.com/science/article/pii/S0003491605002630}
  {\bibfield  {journal} {\bibinfo  {journal} {Ann. Phys.}\ }\textbf {\bibinfo
  {volume} {321}},\ \bibinfo {pages} {1126 } (\bibinfo {year}
  {2006})}\BibitemShut {NoStop}%
\bibitem [{\citenamefont {Gornyi}\ \emph {et~al.}(2005)\citenamefont {Gornyi},
  \citenamefont {Mirlin},\ and\ \citenamefont {Polyakov}}]{Gornyi2005}%
  \BibitemOpen
  \bibfield  {author} {\bibinfo {author} {\bibfnamefont {I.~V.}\ \bibnamefont
  {Gornyi}}, \bibinfo {author} {\bibfnamefont {A.~D.}\ \bibnamefont {Mirlin}},
  \ and\ \bibinfo {author} {\bibfnamefont {D.~G.}\ \bibnamefont {Polyakov}},\
  }\href {\doibase 10.1103/PhysRevLett.95.206603} {\bibfield  {journal}
  {\bibinfo  {journal} {Phys. Rev. Lett.}\ }\textbf {\bibinfo {volume} {95}},\
  \bibinfo {pages} {206603} (\bibinfo {year} {2005})}\BibitemShut {NoStop}%
\bibitem [{\citenamefont {\ifmmode \check{Z}\else
  \v{Z}\fi{}nidari\ifmmode~\check{c}\else \v{c}\fi{}}\ \emph
  {et~al.}(2008)\citenamefont {\ifmmode \check{Z}\else
  \v{Z}\fi{}nidari\ifmmode~\check{c}\else \v{c}\fi{}}, \citenamefont {Prosen},\
  and\ \citenamefont {Prelov\ifmmode~\check{s}\else
  \v{s}\fi{}ek}}]{Znidaric2008}%
  \BibitemOpen
  \bibfield  {author} {\bibinfo {author} {\bibfnamefont {M.}~\bibnamefont
  {\ifmmode \check{Z}\else \v{Z}\fi{}nidari\ifmmode~\check{c}\else
  \v{c}\fi{}}}, \bibinfo {author} {\bibfnamefont {T.}~\bibnamefont {Prosen}}, \
  and\ \bibinfo {author} {\bibfnamefont {P.}~\bibnamefont
  {Prelov\ifmmode~\check{s}\else \v{s}\fi{}ek}},\ }\href {\doibase
  10.1103/PhysRevB.77.064426} {\bibfield  {journal} {\bibinfo  {journal} {Phys.
  Rev. B}\ }\textbf {\bibinfo {volume} {77}},\ \bibinfo {pages} {064426}
  (\bibinfo {year} {2008})}\BibitemShut {NoStop}%
\bibitem [{\citenamefont {Pal}\ and\ \citenamefont {Huse}(2010)}]{Pal2010}%
  \BibitemOpen
  \bibfield  {author} {\bibinfo {author} {\bibfnamefont {A.}~\bibnamefont
  {Pal}}\ and\ \bibinfo {author} {\bibfnamefont {D.~A.}\ \bibnamefont {Huse}},\
  }\href {\doibase 10.1103/PhysRevB.82.174411} {\bibfield  {journal} {\bibinfo
  {journal} {Phys. Rev. B}\ }\textbf {\bibinfo {volume} {82}},\ \bibinfo
  {pages} {174411} (\bibinfo {year} {2010})}\BibitemShut {NoStop}%
\bibitem [{\citenamefont {Bardarson}\ \emph {et~al.}(2012)\citenamefont
  {Bardarson}, \citenamefont {Pollmann},\ and\ \citenamefont
  {Moore}}]{Bardarson2012}%
  \BibitemOpen
  \bibfield  {author} {\bibinfo {author} {\bibfnamefont {J.~H.}\ \bibnamefont
  {Bardarson}}, \bibinfo {author} {\bibfnamefont {F.}~\bibnamefont {Pollmann}},
  \ and\ \bibinfo {author} {\bibfnamefont {J.~E.}\ \bibnamefont {Moore}},\
  }\href {\doibase 10.1103/PhysRevLett.109.017202} {\bibfield  {journal}
  {\bibinfo  {journal} {Phys. Rev. Lett.}\ }\textbf {\bibinfo {volume} {109}},\
  \bibinfo {pages} {017202} (\bibinfo {year} {2012})}\BibitemShut {NoStop}%
\bibitem [{\citenamefont {Nandkishore}\ and\ \citenamefont
  {Huse}(2015)}]{Nandkishore2015}%
  \BibitemOpen
  \bibfield  {author} {\bibinfo {author} {\bibfnamefont {R.}~\bibnamefont
  {Nandkishore}}\ and\ \bibinfo {author} {\bibfnamefont {D.~A.}\ \bibnamefont
  {Huse}},\ }\href {\doibase 10.1146/annurev-conmatphys-031214-014726}
  {\bibfield  {journal} {\bibinfo  {journal} {Ann. Rev. Condens. Matter Phys.}\
  }\textbf {\bibinfo {volume} {6}},\ \bibinfo {pages} {15} (\bibinfo {year}
  {2015})}\BibitemShut {NoStop}%
\bibitem [{\citenamefont {Altman}\ and\ \citenamefont
  {Vosk}(2015)}]{Altman2015}%
  \BibitemOpen
  \bibfield  {author} {\bibinfo {author} {\bibfnamefont {E.}~\bibnamefont
  {Altman}}\ and\ \bibinfo {author} {\bibfnamefont {R.}~\bibnamefont {Vosk}},\
  }\href {\doibase 10.1146/annurev-conmatphys-031214-014701} {\bibfield
  {journal} {\bibinfo  {journal} {Ann. Rev. Condens. Matter Phys.}\ }\textbf
  {\bibinfo {volume} {6}},\ \bibinfo {pages} {383} (\bibinfo {year}
  {2015})}\BibitemShut {NoStop}%
\bibitem [{\citenamefont {Luitz}\ \emph {et~al.}(2015)\citenamefont {Luitz},
  \citenamefont {Laflorencie},\ and\ \citenamefont {Alet}}]{Luitz2015}%
  \BibitemOpen
  \bibfield  {author} {\bibinfo {author} {\bibfnamefont {D.~J.}\ \bibnamefont
  {Luitz}}, \bibinfo {author} {\bibfnamefont {N.}~\bibnamefont {Laflorencie}},
  \ and\ \bibinfo {author} {\bibfnamefont {F.}~\bibnamefont {Alet}},\ }\href
  {\doibase 10.1103/PhysRevB.91.081103} {\bibfield  {journal} {\bibinfo
  {journal} {Phys. Rev. B}\ }\textbf {\bibinfo {volume} {91}},\ \bibinfo
  {pages} {081103} (\bibinfo {year} {2015})}\BibitemShut {NoStop}%
\bibitem [{\citenamefont {Bera}\ \emph {et~al.}(2015)\citenamefont {Bera},
  \citenamefont {Schomerus}, \citenamefont {Heidrich-Meisner},\ and\
  \citenamefont {Bardarson}}]{Bera2015}%
  \BibitemOpen
  \bibfield  {author} {\bibinfo {author} {\bibfnamefont {S.}~\bibnamefont
  {Bera}}, \bibinfo {author} {\bibfnamefont {H.}~\bibnamefont {Schomerus}},
  \bibinfo {author} {\bibfnamefont {F.}~\bibnamefont {Heidrich-Meisner}}, \
  and\ \bibinfo {author} {\bibfnamefont {J.~H.}\ \bibnamefont {Bardarson}},\
  }\href {\doibase 10.1103/PhysRevLett.115.046603} {\bibfield  {journal}
  {\bibinfo  {journal} {Phys. Rev. Lett.}\ }\textbf {\bibinfo {volume} {115}},\
  \bibinfo {pages} {046603} (\bibinfo {year} {2015})}\BibitemShut {NoStop}%
\bibitem [{\citenamefont {\ifmmode \check{Z}\else
  \v{Z}\fi{}nidari\ifmmode~\check{c}\else \v{c}\fi{}}\ \emph
  {et~al.}(2016)\citenamefont {\ifmmode \check{Z}\else
  \v{Z}\fi{}nidari\ifmmode~\check{c}\else \v{c}\fi{}}, \citenamefont
  {Scardicchio},\ and\ \citenamefont {Varma}}]{Znidaric2016}%
  \BibitemOpen
  \bibfield  {author} {\bibinfo {author} {\bibfnamefont {M.}~\bibnamefont
  {\ifmmode \check{Z}\else \v{Z}\fi{}nidari\ifmmode~\check{c}\else
  \v{c}\fi{}}}, \bibinfo {author} {\bibfnamefont {A.}~\bibnamefont
  {Scardicchio}}, \ and\ \bibinfo {author} {\bibfnamefont {V.~K.}\ \bibnamefont
  {Varma}},\ }\href {\doibase 10.1103/PhysRevLett.117.040601} {\bibfield
  {journal} {\bibinfo  {journal} {Phys. Rev. Lett.}\ }\textbf {\bibinfo
  {volume} {117}},\ \bibinfo {pages} {040601} (\bibinfo {year}
  {2016})}\BibitemShut {NoStop}%
\bibitem [{\citenamefont {{Ann. Phys.~(Berlin)}}(2017)}]{AnnMBLReview2017}%
  \BibitemOpen
  \bibfield  {author} {\bibinfo {author} {\bibnamefont {{Ann.
  Phys.~(Berlin)}}}\ }(\bibinfo {year} {2017})\ p.\ \bibinfo {pages}
  {1770050}\BibitemShut {NoStop}%
\bibitem [{\citenamefont {Alet}\ and\ \citenamefont
  {Laflorencie}(2018)}]{Alet2018}%
  \BibitemOpen
  \bibfield  {author} {\bibinfo {author} {\bibfnamefont {F.}~\bibnamefont
  {Alet}}\ and\ \bibinfo {author} {\bibfnamefont {N.}~\bibnamefont
  {Laflorencie}},\ }\href
  {http://www.sciencedirect.com/science/article/pii/S163107051830032X}
  {\bibfield  {journal} {\bibinfo  {journal} {C. R. Phys.}\ } (\bibinfo {year}
  {2018})}\BibitemShut {NoStop}%
\bibitem [{\citenamefont {{Abanin}}\ \emph {et~al.}(2018)\citenamefont
  {{Abanin}}, \citenamefont {{Altman}}, \citenamefont {{Bloch}},\ and\
  \citenamefont {{Serbyn}}}]{AbaninBloch-Review-2018}%
  \BibitemOpen
  \bibfield  {author} {\bibinfo {author} {\bibfnamefont {D.~A.}\ \bibnamefont
  {{Abanin}}}, \bibinfo {author} {\bibfnamefont {E.}~\bibnamefont {{Altman}}},
  \bibinfo {author} {\bibfnamefont {I.}~\bibnamefont {{Bloch}}}, \ and\
  \bibinfo {author} {\bibfnamefont {M.}~\bibnamefont {{Serbyn}}},\ }\href@noop
  {} {\bibfield  {journal} {\bibinfo  {journal} {arXiv e-prints}\ ,\ \bibinfo
  {eid} {arXiv:1804.11065}} (\bibinfo {year} {2018})},\ \Eprint
  {http://arxiv.org/abs/1804.11065} {arXiv:1804.11065 [cond-mat.dis-nn]}
  \BibitemShut {NoStop}%
\bibitem [{\citenamefont {Schreiber}\ \emph {et~al.}(2015)\citenamefont
  {Schreiber}, \citenamefont {Hodgman}, \citenamefont {Bordia}, \citenamefont
  {L{\"u}schen}, \citenamefont {Fischer}, \citenamefont {Vosk}, \citenamefont
  {Altman}, \citenamefont {Schneider},\ and\ \citenamefont
  {Bloch}}]{Schreiber2015}%
  \BibitemOpen
  \bibfield  {author} {\bibinfo {author} {\bibfnamefont {M.}~\bibnamefont
  {Schreiber}}, \bibinfo {author} {\bibfnamefont {S.~S.}\ \bibnamefont
  {Hodgman}}, \bibinfo {author} {\bibfnamefont {P.}~\bibnamefont {Bordia}},
  \bibinfo {author} {\bibfnamefont {H.~P.}\ \bibnamefont {L{\"u}schen}},
  \bibinfo {author} {\bibfnamefont {M.~H.}\ \bibnamefont {Fischer}}, \bibinfo
  {author} {\bibfnamefont {R.}~\bibnamefont {Vosk}}, \bibinfo {author}
  {\bibfnamefont {E.}~\bibnamefont {Altman}}, \bibinfo {author} {\bibfnamefont
  {U.}~\bibnamefont {Schneider}}, \ and\ \bibinfo {author} {\bibfnamefont
  {I.}~\bibnamefont {Bloch}},\ }\href {\doibase 10.1126/science.aaa7432}
  {\bibfield  {journal} {\bibinfo  {journal} {Science}\ }\textbf {\bibinfo
  {volume} {349}},\ \bibinfo {pages} {842} (\bibinfo {year}
  {2015})}\BibitemShut {NoStop}%
\bibitem [{\citenamefont {Choi}\ \emph {et~al.}(2016)\citenamefont {Choi},
  \citenamefont {Hild}, \citenamefont {Zeiher}, \citenamefont {Schau{\ss}},
  \citenamefont {Rubio-Abadal}, \citenamefont {Yefsah}, \citenamefont
  {Khemani}, \citenamefont {Huse}, \citenamefont {Bloch},\ and\ \citenamefont
  {Gross}}]{Choi2016}%
  \BibitemOpen
  \bibfield  {author} {\bibinfo {author} {\bibfnamefont {J.-Y.}\ \bibnamefont
  {Choi}}, \bibinfo {author} {\bibfnamefont {S.}~\bibnamefont {Hild}}, \bibinfo
  {author} {\bibfnamefont {J.}~\bibnamefont {Zeiher}}, \bibinfo {author}
  {\bibfnamefont {P.}~\bibnamefont {Schau{\ss}}}, \bibinfo {author}
  {\bibfnamefont {A.}~\bibnamefont {Rubio-Abadal}}, \bibinfo {author}
  {\bibfnamefont {T.}~\bibnamefont {Yefsah}}, \bibinfo {author} {\bibfnamefont
  {V.}~\bibnamefont {Khemani}}, \bibinfo {author} {\bibfnamefont {D.~A.}\
  \bibnamefont {Huse}}, \bibinfo {author} {\bibfnamefont {I.}~\bibnamefont
  {Bloch}}, \ and\ \bibinfo {author} {\bibfnamefont {C.}~\bibnamefont
  {Gross}},\ }\href {\doibase 10.1126/science.aaf8834} {\bibfield  {journal}
  {\bibinfo  {journal} {Science}\ }\textbf {\bibinfo {volume} {352}},\ \bibinfo
  {pages} {1547} (\bibinfo {year} {2016})}\BibitemShut {NoStop}%
\bibitem [{\citenamefont {L\"uschen}\ \emph {et~al.}(2017)\citenamefont
  {L\"uschen}, \citenamefont {Bordia}, \citenamefont {Scherg}, \citenamefont
  {Alet}, \citenamefont {Altman}, \citenamefont {Schneider},\ and\
  \citenamefont {Bloch}}]{Luschen2017}%
  \BibitemOpen
  \bibfield  {author} {\bibinfo {author} {\bibfnamefont {H.~P.}\ \bibnamefont
  {L\"uschen}}, \bibinfo {author} {\bibfnamefont {P.}~\bibnamefont {Bordia}},
  \bibinfo {author} {\bibfnamefont {S.}~\bibnamefont {Scherg}}, \bibinfo
  {author} {\bibfnamefont {F.}~\bibnamefont {Alet}}, \bibinfo {author}
  {\bibfnamefont {E.}~\bibnamefont {Altman}}, \bibinfo {author} {\bibfnamefont
  {U.}~\bibnamefont {Schneider}}, \ and\ \bibinfo {author} {\bibfnamefont
  {I.}~\bibnamefont {Bloch}},\ }\href {\doibase 10.1103/PhysRevLett.119.260401}
  {\bibfield  {journal} {\bibinfo  {journal} {Phys. Rev. Lett.}\ }\textbf
  {\bibinfo {volume} {119}},\ \bibinfo {pages} {260401} (\bibinfo {year}
  {2017})}\BibitemShut {NoStop}%
\bibitem [{\citenamefont {Bordia}\ \emph {et~al.}(2017)\citenamefont {Bordia},
  \citenamefont {L\"uschen}, \citenamefont {Scherg}, \citenamefont
  {Gopalakrishnan}, \citenamefont {Knap}, \citenamefont {Schneider},\ and\
  \citenamefont {Bloch}}]{BordiaPRX2017}%
  \BibitemOpen
  \bibfield  {author} {\bibinfo {author} {\bibfnamefont {P.}~\bibnamefont
  {Bordia}}, \bibinfo {author} {\bibfnamefont {H.}~\bibnamefont {L\"uschen}},
  \bibinfo {author} {\bibfnamefont {S.}~\bibnamefont {Scherg}}, \bibinfo
  {author} {\bibfnamefont {S.}~\bibnamefont {Gopalakrishnan}}, \bibinfo
  {author} {\bibfnamefont {M.}~\bibnamefont {Knap}}, \bibinfo {author}
  {\bibfnamefont {U.}~\bibnamefont {Schneider}}, \ and\ \bibinfo {author}
  {\bibfnamefont {I.}~\bibnamefont {Bloch}},\ }\href {\doibase
  10.1103/PhysRevX.7.041047} {\bibfield  {journal} {\bibinfo  {journal} {Phys.
  Rev. X}\ }\textbf {\bibinfo {volume} {7}},\ \bibinfo {pages} {041047}
  (\bibinfo {year} {2017})}\BibitemShut {NoStop}%
\bibitem [{\citenamefont {{Kohlert}}\ \emph {et~al.}(2018)\citenamefont
  {{Kohlert}}, \citenamefont {{Scherg}}, \citenamefont {{Li}}, \citenamefont
  {{L{\"u}schen}}, \citenamefont {{Das Sarma}}, \citenamefont {{Bloch}},\ and\
  \citenamefont {{Aidelsburger}}}]{Kohlert2018}%
  \BibitemOpen
  \bibfield  {author} {\bibinfo {author} {\bibfnamefont {T.}~\bibnamefont
  {{Kohlert}}}, \bibinfo {author} {\bibfnamefont {S.}~\bibnamefont {{Scherg}}},
  \bibinfo {author} {\bibfnamefont {X.}~\bibnamefont {{Li}}}, \bibinfo {author}
  {\bibfnamefont {H.~P.}\ \bibnamefont {{L{\"u}schen}}}, \bibinfo {author}
  {\bibfnamefont {S.}~\bibnamefont {{Das Sarma}}}, \bibinfo {author}
  {\bibfnamefont {I.}~\bibnamefont {{Bloch}}}, \ and\ \bibinfo {author}
  {\bibfnamefont {M.}~\bibnamefont {{Aidelsburger}}},\ }\href@noop {}
  {\bibfield  {journal} {\bibinfo  {journal} {arXiv e-prints}\ ,\ \bibinfo
  {eid} {arXiv:1809.04055}} (\bibinfo {year} {2018})},\ \Eprint
  {http://arxiv.org/abs/1809.04055} {arXiv:1809.04055 [cond-mat.quant-gas]}
  \BibitemShut {NoStop}%
\bibitem [{\citenamefont {{Rispoli}}\ \emph {et~al.}(2018)\citenamefont
  {{Rispoli}}, \citenamefont {{Lukin}}, \citenamefont {{Schittko}},
  \citenamefont {{Kim}}, \citenamefont {{Tai}}, \citenamefont {{L{\'e}onard}},\
  and\ \citenamefont {{Greiner}}}]{RispoliExp18}%
  \BibitemOpen
  \bibfield  {author} {\bibinfo {author} {\bibfnamefont {M.}~\bibnamefont
  {{Rispoli}}}, \bibinfo {author} {\bibfnamefont {A.}~\bibnamefont {{Lukin}}},
  \bibinfo {author} {\bibfnamefont {R.}~\bibnamefont {{Schittko}}}, \bibinfo
  {author} {\bibfnamefont {S.}~\bibnamefont {{Kim}}}, \bibinfo {author}
  {\bibfnamefont {M.~E.}\ \bibnamefont {{Tai}}}, \bibinfo {author}
  {\bibfnamefont {J.}~\bibnamefont {{L{\'e}onard}}}, \ and\ \bibinfo {author}
  {\bibfnamefont {M.}~\bibnamefont {{Greiner}}},\ }\href@noop {} {\bibfield
  {journal} {\bibinfo  {journal} {arXiv e-prints}\ ,\ \bibinfo {eid}
  {arXiv:1812.06959}} (\bibinfo {year} {2018})},\ \Eprint
  {http://arxiv.org/abs/1812.06959} {arXiv:1812.06959 [cond-mat.quant-gas]}
  \BibitemShut {NoStop}%
\bibitem [{\citenamefont {Wei}\ \emph {et~al.}(2018)\citenamefont {Wei},
  \citenamefont {Ramanathan},\ and\ \citenamefont
  {Cappellaro}}]{WeiNuclearSpin18}%
  \BibitemOpen
  \bibfield  {author} {\bibinfo {author} {\bibfnamefont {K.~X.}\ \bibnamefont
  {Wei}}, \bibinfo {author} {\bibfnamefont {C.}~\bibnamefont {Ramanathan}}, \
  and\ \bibinfo {author} {\bibfnamefont {P.}~\bibnamefont {Cappellaro}},\
  }\href {\doibase 10.1103/PhysRevLett.120.070501} {\bibfield  {journal}
  {\bibinfo  {journal} {Phys. Rev. Lett.}\ }\textbf {\bibinfo {volume} {120}},\
  \bibinfo {pages} {070501} (\bibinfo {year} {2018})}\BibitemShut {NoStop}%
\bibitem [{\citenamefont {Smith}\ \emph {et~al.}(2016)\citenamefont {Smith},
  \citenamefont {Lee}, \citenamefont {Richerme}, \citenamefont {Neyenhuis},
  \citenamefont {Hess}, \citenamefont {Hauke}, \citenamefont {Heyl},
  \citenamefont {Huse},\ and\ \citenamefont {Monroe}}]{Smith2016}%
  \BibitemOpen
  \bibfield  {author} {\bibinfo {author} {\bibfnamefont {J.}~\bibnamefont
  {Smith}}, \bibinfo {author} {\bibfnamefont {A.}~\bibnamefont {Lee}}, \bibinfo
  {author} {\bibfnamefont {P.}~\bibnamefont {Richerme}}, \bibinfo {author}
  {\bibfnamefont {B.}~\bibnamefont {Neyenhuis}}, \bibinfo {author}
  {\bibfnamefont {P.~W.}\ \bibnamefont {Hess}}, \bibinfo {author}
  {\bibfnamefont {P.}~\bibnamefont {Hauke}}, \bibinfo {author} {\bibfnamefont
  {M.}~\bibnamefont {Heyl}}, \bibinfo {author} {\bibfnamefont {D.~A.}\
  \bibnamefont {Huse}}, \ and\ \bibinfo {author} {\bibfnamefont
  {C.}~\bibnamefont {Monroe}},\ }\href {http://dx.doi.org/10.1038/nphys3783}
  {\bibfield  {journal} {\bibinfo  {journal} {Nat. Phys.}\ }\textbf {\bibinfo
  {volume} {advance online publication}} (\bibinfo {year} {2016})}\BibitemShut
  {NoStop}%
\bibitem [{\citenamefont {Kucsko}\ \emph {et~al.}(2018)\citenamefont {Kucsko},
  \citenamefont {Choi}, \citenamefont {Choi}, \citenamefont {Maurer},
  \citenamefont {Zhou}, \citenamefont {Landig}, \citenamefont {Sumiya},
  \citenamefont {Onoda}, \citenamefont {Isoya}, \citenamefont {Jelezko},
  \citenamefont {Demler}, \citenamefont {Yao},\ and\ \citenamefont
  {Lukin}}]{KucskoDiamondPRL2018}%
  \BibitemOpen
  \bibfield  {author} {\bibinfo {author} {\bibfnamefont {G.}~\bibnamefont
  {Kucsko}}, \bibinfo {author} {\bibfnamefont {S.}~\bibnamefont {Choi}},
  \bibinfo {author} {\bibfnamefont {J.}~\bibnamefont {Choi}}, \bibinfo {author}
  {\bibfnamefont {P.~C.}\ \bibnamefont {Maurer}}, \bibinfo {author}
  {\bibfnamefont {H.}~\bibnamefont {Zhou}}, \bibinfo {author} {\bibfnamefont
  {R.}~\bibnamefont {Landig}}, \bibinfo {author} {\bibfnamefont
  {H.}~\bibnamefont {Sumiya}}, \bibinfo {author} {\bibfnamefont
  {S.}~\bibnamefont {Onoda}}, \bibinfo {author} {\bibfnamefont
  {J.}~\bibnamefont {Isoya}}, \bibinfo {author} {\bibfnamefont
  {F.}~\bibnamefont {Jelezko}}, \bibinfo {author} {\bibfnamefont
  {E.}~\bibnamefont {Demler}}, \bibinfo {author} {\bibfnamefont {N.~Y.}\
  \bibnamefont {Yao}}, \ and\ \bibinfo {author} {\bibfnamefont {M.~D.}\
  \bibnamefont {Lukin}},\ }\href {\doibase 10.1103/PhysRevLett.121.023601}
  {\bibfield  {journal} {\bibinfo  {journal} {Phys. Rev. Lett.}\ }\textbf
  {\bibinfo {volume} {121}},\ \bibinfo {pages} {023601} (\bibinfo {year}
  {2018})}\BibitemShut {NoStop}%
\bibitem [{\citenamefont {Ovadia}\ \emph {et~al.}(2015)\citenamefont {Ovadia},
  \citenamefont {Kalok}, \citenamefont {Tamir}, \citenamefont {Mitra},
  \citenamefont {Sac{\'e}p{\'e}},\ and\ \citenamefont {Shahar}}]{Ovadia2015}%
  \BibitemOpen
  \bibfield  {author} {\bibinfo {author} {\bibfnamefont {M.}~\bibnamefont
  {Ovadia}}, \bibinfo {author} {\bibfnamefont {D.}~\bibnamefont {Kalok}},
  \bibinfo {author} {\bibfnamefont {I.}~\bibnamefont {Tamir}}, \bibinfo
  {author} {\bibfnamefont {S.}~\bibnamefont {Mitra}}, \bibinfo {author}
  {\bibfnamefont {B.}~\bibnamefont {Sac{\'e}p{\'e}}}, \ and\ \bibinfo {author}
  {\bibfnamefont {D.}~\bibnamefont {Shahar}},\ }\href
  {http://dx.doi.org/10.1038/srep13503} {\bibfield  {journal} {\bibinfo
  {journal} {Sci. Rep.}\ }\textbf {\bibinfo {volume} {5}},\ \bibinfo {pages}
  {13503 EP} (\bibinfo {year} {2015})}\BibitemShut {NoStop}%
\bibitem [{\citenamefont {Bera}\ \emph {et~al.}(2017)\citenamefont {Bera},
  \citenamefont {De~Tomasi}, \citenamefont {Weiner},\ and\ \citenamefont
  {Evers}}]{Bera2017}%
  \BibitemOpen
  \bibfield  {author} {\bibinfo {author} {\bibfnamefont {S.}~\bibnamefont
  {Bera}}, \bibinfo {author} {\bibfnamefont {G.}~\bibnamefont {De~Tomasi}},
  \bibinfo {author} {\bibfnamefont {F.}~\bibnamefont {Weiner}}, \ and\ \bibinfo
  {author} {\bibfnamefont {F.}~\bibnamefont {Evers}},\ }\href {\doibase
  10.1103/PhysRevLett.118.196801} {\bibfield  {journal} {\bibinfo  {journal}
  {Phys. Rev. Lett.}\ }\textbf {\bibinfo {volume} {118}},\ \bibinfo {pages}
  {196801} (\bibinfo {year} {2017})}\BibitemShut {NoStop}%
\bibitem [{Note1()}]{Note1}%
  \BibitemOpen
  \bibinfo {note} {For example,~\protect \rev@citet {Luitz2016} report values
  for the imbalance exponent $\zeta $ that are smaller than the values detected
  by~\protect \rev@citet {Doggen2018} by a factor of two or more, e.g., near
  $W{=}2$. Both works have tested convergence with respect to numerical
  parameters, while they differ in the choice of the initial state:
  random~\cite {Luitz2016} versus N\'eel~\cite {Doggen2018}.}\BibitemShut
  {Stop}%
\bibitem [{\citenamefont {Vosk}\ \emph {et~al.}(2015)\citenamefont {Vosk},
  \citenamefont {Huse},\ and\ \citenamefont {Altman}}]{Vosk2015}%
  \BibitemOpen
  \bibfield  {author} {\bibinfo {author} {\bibfnamefont {R.}~\bibnamefont
  {Vosk}}, \bibinfo {author} {\bibfnamefont {D.~A.}\ \bibnamefont {Huse}}, \
  and\ \bibinfo {author} {\bibfnamefont {E.}~\bibnamefont {Altman}},\ }\href
  {\doibase 10.1103/PhysRevX.5.031032} {\bibfield  {journal} {\bibinfo
  {journal} {Phys. Rev. X}\ }\textbf {\bibinfo {volume} {5}},\ \bibinfo {pages}
  {031032} (\bibinfo {year} {2015})}\BibitemShut {NoStop}%
\bibitem [{\citenamefont {Potter}\ \emph {et~al.}(2015)\citenamefont {Potter},
  \citenamefont {Vasseur},\ and\ \citenamefont {Parameswaran}}]{Potter2015}%
  \BibitemOpen
  \bibfield  {author} {\bibinfo {author} {\bibfnamefont {A.~C.}\ \bibnamefont
  {Potter}}, \bibinfo {author} {\bibfnamefont {R.}~\bibnamefont {Vasseur}}, \
  and\ \bibinfo {author} {\bibfnamefont {S.~A.}\ \bibnamefont {Parameswaran}},\
  }\href {\doibase 10.1103/PhysRevX.5.031033} {\bibfield  {journal} {\bibinfo
  {journal} {Phys. Rev. X}\ }\textbf {\bibinfo {volume} {5}},\ \bibinfo {pages}
  {031033} (\bibinfo {year} {2015})}\BibitemShut {NoStop}%
\bibitem [{\citenamefont {Agarwal}\ \emph {et~al.}(2017)\citenamefont
  {Agarwal}, \citenamefont {Altman}, \citenamefont {Demler}, \citenamefont
  {Gopalakrishnan}, \citenamefont {Huse},\ and\ \citenamefont
  {Knap}}]{Agarwal2017}%
  \BibitemOpen
  \bibfield  {author} {\bibinfo {author} {\bibfnamefont {K.}~\bibnamefont
  {Agarwal}}, \bibinfo {author} {\bibfnamefont {E.}~\bibnamefont {Altman}},
  \bibinfo {author} {\bibfnamefont {E.}~\bibnamefont {Demler}}, \bibinfo
  {author} {\bibfnamefont {S.}~\bibnamefont {Gopalakrishnan}}, \bibinfo
  {author} {\bibfnamefont {D.~A.}\ \bibnamefont {Huse}}, \ and\ \bibinfo
  {author} {\bibfnamefont {M.}~\bibnamefont {Knap}},\ }\href
  {https://onlinelibrary.wiley.com/doi/abs/10.1002/andp.201600326} {\bibfield
  {journal} {\bibinfo  {journal} {Ann. Phys. (Berlin)}\ }\textbf {\bibinfo
  {volume} {529}},\ \bibinfo {pages} {1600326} (\bibinfo {year}
  {2017})}\BibitemShut {NoStop}%
\bibitem [{\citenamefont {Luitz}\ and\ \citenamefont
  {Lev}(2017)}]{Luitz:2017cp}%
  \BibitemOpen
  \bibfield  {author} {\bibinfo {author} {\bibfnamefont {D.~J.}\ \bibnamefont
  {Luitz}}\ and\ \bibinfo {author} {\bibfnamefont {Y.~B.}\ \bibnamefont
  {Lev}},\ }\href
  {https://onlinelibrary.wiley.com/doi/abs/10.1002/andp.201600350} {\bibfield
  {journal} {\bibinfo  {journal} {Ann. Phys. (Berlin)}\ }\textbf {\bibinfo
  {volume} {529}},\ \bibinfo {pages} {1600350} (\bibinfo {year}
  {2017})}\BibitemShut {NoStop}%
\bibitem [{\citenamefont {Prelovšek}\ \emph {et~al.}(2017)\citenamefont
  {Prelovšek}, \citenamefont {Mierzejewski}, \citenamefont {Barišić},\ and\
  \citenamefont {Herbrych}}]{Prelovsek-review-2017}%
  \BibitemOpen
  \bibfield  {author} {\bibinfo {author} {\bibfnamefont {P.}~\bibnamefont
  {Prelovšek}}, \bibinfo {author} {\bibfnamefont {M.}~\bibnamefont
  {Mierzejewski}}, \bibinfo {author} {\bibfnamefont {O.}~\bibnamefont
  {Barišić}}, \ and\ \bibinfo {author} {\bibfnamefont {J.}~\bibnamefont
  {Herbrych}},\ }\href {\doibase 10.1002/andp.201600362} {\bibfield  {journal}
  {\bibinfo  {journal} {Ann. Phys. (Berlin)}\ }\textbf {\bibinfo {volume}
  {529}},\ \bibinfo {pages} {1600362} (\bibinfo {year} {2017})}\BibitemShut
  {NoStop}%
\bibitem [{\citenamefont {Iyer}\ \emph {et~al.}(2013)\citenamefont {Iyer},
  \citenamefont {Oganesyan}, \citenamefont {Refael},\ and\ \citenamefont
  {Huse}}]{Iyer13}%
  \BibitemOpen
  \bibfield  {author} {\bibinfo {author} {\bibfnamefont {S.}~\bibnamefont
  {Iyer}}, \bibinfo {author} {\bibfnamefont {V.}~\bibnamefont {Oganesyan}},
  \bibinfo {author} {\bibfnamefont {G.}~\bibnamefont {Refael}}, \ and\ \bibinfo
  {author} {\bibfnamefont {D.~A.}\ \bibnamefont {Huse}},\ }\href {\doibase
  10.1103/PhysRevB.87.134202} {\bibfield  {journal} {\bibinfo  {journal} {Phys.
  Rev. B}\ }\textbf {\bibinfo {volume} {87}},\ \bibinfo {pages} {134202}
  (\bibinfo {year} {2013})}\BibitemShut {NoStop}%
\bibitem [{\citenamefont {Lee}\ \emph {et~al.}(2017)\citenamefont {Lee},
  \citenamefont {Look}, \citenamefont {Lim},\ and\ \citenamefont
  {Sheng}}]{Lee17}%
  \BibitemOpen
  \bibfield  {author} {\bibinfo {author} {\bibfnamefont {M.}~\bibnamefont
  {Lee}}, \bibinfo {author} {\bibfnamefont {T.~R.}\ \bibnamefont {Look}},
  \bibinfo {author} {\bibfnamefont {S.~P.}\ \bibnamefont {Lim}}, \ and\
  \bibinfo {author} {\bibfnamefont {D.~N.}\ \bibnamefont {Sheng}},\ }\href
  {\doibase 10.1103/PhysRevB.96.075146} {\bibfield  {journal} {\bibinfo
  {journal} {Phys. Rev. B}\ }\textbf {\bibinfo {volume} {96}},\ \bibinfo
  {pages} {075146} (\bibinfo {year} {2017})}\BibitemShut {NoStop}%
\bibitem [{\citenamefont {Khemani}\ \emph
  {et~al.}(2017{\natexlab{a}})\citenamefont {Khemani}, \citenamefont {Sheng},\
  and\ \citenamefont {Huse}}]{KhemaniPRL17}%
  \BibitemOpen
  \bibfield  {author} {\bibinfo {author} {\bibfnamefont {V.}~\bibnamefont
  {Khemani}}, \bibinfo {author} {\bibfnamefont {D.~N.}\ \bibnamefont {Sheng}},
  \ and\ \bibinfo {author} {\bibfnamefont {D.~A.}\ \bibnamefont {Huse}},\
  }\href {\doibase 10.1103/PhysRevLett.119.075702} {\bibfield  {journal}
  {\bibinfo  {journal} {Phys. Rev. Lett.}\ }\textbf {\bibinfo {volume} {119}},\
  \bibinfo {pages} {075702} (\bibinfo {year} {2017}{\natexlab{a}})}\BibitemShut
  {NoStop}%
\bibitem [{\citenamefont {Setiawan}\ \emph {et~al.}(2017)\citenamefont
  {Setiawan}, \citenamefont {Deng},\ and\ \citenamefont
  {Pixley}}]{Setiawan2017}%
  \BibitemOpen
  \bibfield  {author} {\bibinfo {author} {\bibfnamefont {F.}~\bibnamefont
  {Setiawan}}, \bibinfo {author} {\bibfnamefont {D.-L.}\ \bibnamefont {Deng}},
  \ and\ \bibinfo {author} {\bibfnamefont {J.~H.}\ \bibnamefont {Pixley}},\
  }\href {\doibase 10.1103/PhysRevB.96.104205} {\bibfield  {journal} {\bibinfo
  {journal} {Phys. Rev. B}\ }\textbf {\bibinfo {volume} {96}},\ \bibinfo
  {pages} {104205} (\bibinfo {year} {2017})}\BibitemShut {NoStop}%
\bibitem [{\citenamefont {{Doggen}}\ and\ \citenamefont
  {{Mirlin}}(2019)}]{Doggen2019}%
  \BibitemOpen
  \bibfield  {author} {\bibinfo {author} {\bibfnamefont {E.~V.~H.}\
  \bibnamefont {{Doggen}}}\ and\ \bibinfo {author} {\bibfnamefont {A.~D.}\
  \bibnamefont {{Mirlin}}},\ }\href@noop {} {\bibfield  {journal} {\bibinfo
  {journal} {arXiv e-prints}\ ,\ \bibinfo {eid} {arXiv:1901.06971}} (\bibinfo
  {year} {2019})},\ \Eprint {http://arxiv.org/abs/1901.06971} {arXiv:1901.06971
  [cond-mat.dis-nn]} \BibitemShut {NoStop}%
\bibitem [{\citenamefont {Bar~Lev}\ \emph {et~al.}(2015)\citenamefont
  {Bar~Lev}, \citenamefont {Cohen},\ and\ \citenamefont
  {Reichman}}]{BarLevPRL2015}%
  \BibitemOpen
  \bibfield  {author} {\bibinfo {author} {\bibfnamefont {Y.}~\bibnamefont
  {Bar~Lev}}, \bibinfo {author} {\bibfnamefont {G.}~\bibnamefont {Cohen}}, \
  and\ \bibinfo {author} {\bibfnamefont {D.~R.}\ \bibnamefont {Reichman}},\
  }\href {\doibase 10.1103/PhysRevLett.114.100601} {\bibfield  {journal}
  {\bibinfo  {journal} {Phys. Rev. Lett.}\ }\textbf {\bibinfo {volume} {114}},\
  \bibinfo {pages} {100601} (\bibinfo {year} {2015})}\BibitemShut {NoStop}%
\bibitem [{\citenamefont {Lev}\ and\ \citenamefont
  {Reichman}(2016)}]{BarlevEPL2016}%
  \BibitemOpen
  \bibfield  {author} {\bibinfo {author} {\bibfnamefont {Y.~B.}\ \bibnamefont
  {Lev}}\ and\ \bibinfo {author} {\bibfnamefont {D.~R.}\ \bibnamefont
  {Reichman}},\ }\href {http://stacks.iop.org/0295-5075/113/i=4/a=46001}
  {\bibfield  {journal} {\bibinfo  {journal} {EPL (Europhysics Letters)}\
  }\textbf {\bibinfo {volume} {113}},\ \bibinfo {pages} {46001} (\bibinfo
  {year} {2016})}\BibitemShut {NoStop}%
\bibitem [{\citenamefont {Agarwal}\ \emph {et~al.}(2015)\citenamefont
  {Agarwal}, \citenamefont {Gopalakrishnan}, \citenamefont {Knap},
  \citenamefont {M\"uller},\ and\ \citenamefont {Demler}}]{AgarwalPRL2015}%
  \BibitemOpen
  \bibfield  {author} {\bibinfo {author} {\bibfnamefont {K.}~\bibnamefont
  {Agarwal}}, \bibinfo {author} {\bibfnamefont {S.}~\bibnamefont
  {Gopalakrishnan}}, \bibinfo {author} {\bibfnamefont {M.}~\bibnamefont
  {Knap}}, \bibinfo {author} {\bibfnamefont {M.}~\bibnamefont {M\"uller}}, \
  and\ \bibinfo {author} {\bibfnamefont {E.}~\bibnamefont {Demler}},\ }\href
  {\doibase 10.1103/PhysRevLett.114.160401} {\bibfield  {journal} {\bibinfo
  {journal} {Phys. Rev. Lett.}\ }\textbf {\bibinfo {volume} {114}},\ \bibinfo
  {pages} {160401} (\bibinfo {year} {2015})}\BibitemShut {NoStop}%
\bibitem [{\citenamefont {Gopalakrishnan}\ \emph {et~al.}(2015)\citenamefont
  {Gopalakrishnan}, \citenamefont {M\"uller}, \citenamefont {Khemani},
  \citenamefont {Knap}, \citenamefont {Demler},\ and\ \citenamefont
  {Huse}}]{GopalakrishnanPRB15}%
  \BibitemOpen
  \bibfield  {author} {\bibinfo {author} {\bibfnamefont {S.}~\bibnamefont
  {Gopalakrishnan}}, \bibinfo {author} {\bibfnamefont {M.}~\bibnamefont
  {M\"uller}}, \bibinfo {author} {\bibfnamefont {V.}~\bibnamefont {Khemani}},
  \bibinfo {author} {\bibfnamefont {M.}~\bibnamefont {Knap}}, \bibinfo {author}
  {\bibfnamefont {E.}~\bibnamefont {Demler}}, \ and\ \bibinfo {author}
  {\bibfnamefont {D.~A.}\ \bibnamefont {Huse}},\ }\href {\doibase
  10.1103/PhysRevB.92.104202} {\bibfield  {journal} {\bibinfo  {journal} {Phys.
  Rev. B}\ }\textbf {\bibinfo {volume} {92}},\ \bibinfo {pages} {104202}
  (\bibinfo {year} {2015})}\BibitemShut {NoStop}%
\bibitem [{\citenamefont {Khait}\ \emph {et~al.}(2016)\citenamefont {Khait},
  \citenamefont {Gazit}, \citenamefont {Yao},\ and\ \citenamefont
  {Auerbach}}]{KhaitPRB16}%
  \BibitemOpen
  \bibfield  {author} {\bibinfo {author} {\bibfnamefont {I.}~\bibnamefont
  {Khait}}, \bibinfo {author} {\bibfnamefont {S.}~\bibnamefont {Gazit}},
  \bibinfo {author} {\bibfnamefont {N.~Y.}\ \bibnamefont {Yao}}, \ and\
  \bibinfo {author} {\bibfnamefont {A.}~\bibnamefont {Auerbach}},\ }\href
  {\doibase 10.1103/PhysRevB.93.224205} {\bibfield  {journal} {\bibinfo
  {journal} {Phys. Rev. B}\ }\textbf {\bibinfo {volume} {93}},\ \bibinfo
  {pages} {224205} (\bibinfo {year} {2016})}\BibitemShut {NoStop}%
\bibitem [{\citenamefont {Steinigeweg}\ \emph {et~al.}(2016)\citenamefont
  {Steinigeweg}, \citenamefont {Herbrych}, \citenamefont {Pollmann},\ and\
  \citenamefont {Brenig}}]{SteinigewegPRB16}%
  \BibitemOpen
  \bibfield  {author} {\bibinfo {author} {\bibfnamefont {R.}~\bibnamefont
  {Steinigeweg}}, \bibinfo {author} {\bibfnamefont {J.}~\bibnamefont
  {Herbrych}}, \bibinfo {author} {\bibfnamefont {F.}~\bibnamefont {Pollmann}},
  \ and\ \bibinfo {author} {\bibfnamefont {W.}~\bibnamefont {Brenig}},\ }\href
  {\doibase 10.1103/PhysRevB.94.180401} {\bibfield  {journal} {\bibinfo
  {journal} {Phys. Rev. B}\ }\textbf {\bibinfo {volume} {94}},\ \bibinfo
  {pages} {180401} (\bibinfo {year} {2016})}\BibitemShut {NoStop}%
\bibitem [{\citenamefont {Bari\ifmmode \check{s}\else
  \v{s}\fi{}i\ifmmode~\acute{c}\else \'{c}\fi{}}\ \emph
  {et~al.}(2016)\citenamefont {Bari\ifmmode \check{s}\else
  \v{s}\fi{}i\ifmmode~\acute{c}\else \'{c}\fi{}}, \citenamefont {Kokalj},
  \citenamefont {Balog},\ and\ \citenamefont {Prelov\ifmmode~\check{s}\else
  \v{s}\fi{}ek}}]{PrelovsekPRB16}%
  \BibitemOpen
  \bibfield  {author} {\bibinfo {author} {\bibfnamefont {O.~S.}\ \bibnamefont
  {Bari\ifmmode \check{s}\else \v{s}\fi{}i\ifmmode~\acute{c}\else \'{c}\fi{}}},
  \bibinfo {author} {\bibfnamefont {J.}~\bibnamefont {Kokalj}}, \bibinfo
  {author} {\bibfnamefont {I.}~\bibnamefont {Balog}}, \ and\ \bibinfo {author}
  {\bibfnamefont {P.}~\bibnamefont {Prelov\ifmmode~\check{s}\else
  \v{s}\fi{}ek}},\ }\href {\doibase 10.1103/PhysRevB.94.045126} {\bibfield
  {journal} {\bibinfo  {journal} {Phys. Rev. B}\ }\textbf {\bibinfo {volume}
  {94}},\ \bibinfo {pages} {045126} (\bibinfo {year} {2016})}\BibitemShut
  {NoStop}%
\bibitem [{\citenamefont {Doggen}\ \emph {et~al.}(2018)\citenamefont {Doggen},
  \citenamefont {Schindler}, \citenamefont {Tikhonov}, \citenamefont {Mirlin},
  \citenamefont {Neupert}, \citenamefont {Polyakov},\ and\ \citenamefont
  {Gornyi}}]{Doggen2018}%
  \BibitemOpen
  \bibfield  {author} {\bibinfo {author} {\bibfnamefont {E.~V.~H.}\
  \bibnamefont {Doggen}}, \bibinfo {author} {\bibfnamefont {F.}~\bibnamefont
  {Schindler}}, \bibinfo {author} {\bibfnamefont {K.~S.}\ \bibnamefont
  {Tikhonov}}, \bibinfo {author} {\bibfnamefont {A.~D.}\ \bibnamefont
  {Mirlin}}, \bibinfo {author} {\bibfnamefont {T.}~\bibnamefont {Neupert}},
  \bibinfo {author} {\bibfnamefont {D.~G.}\ \bibnamefont {Polyakov}}, \ and\
  \bibinfo {author} {\bibfnamefont {I.~V.}\ \bibnamefont {Gornyi}},\ }\href
  {\doibase 10.1103/PhysRevB.98.174202} {\bibfield  {journal} {\bibinfo
  {journal} {Phys. Rev. B}\ }\textbf {\bibinfo {volume} {98}},\ \bibinfo
  {pages} {174202} (\bibinfo {year} {2018})}\BibitemShut {NoStop}%
\bibitem [{Note2()}]{Note2}%
  \BibitemOpen
  \bibinfo {note} {We include $\protect \mathaccentV {dot}05F\beta (t)$ as an
  indicator with the idea that $t_\xi $ is the only relevant time scale.
  Correspondingly, at $t\gtrsim t_\xi $ the sign of $\beta $ is converged,
  while it numerical value may not be, yet.}\BibitemShut {Stop}%
\bibitem [{\citenamefont {Dumitrescu}\ \emph {et~al.}(2019)\citenamefont
  {Dumitrescu}, \citenamefont {Goremykina}, \citenamefont {Parameswaran},
  \citenamefont {Serbyn},\ and\ \citenamefont {Vasseur}}]{Dumitrescu2018}%
  \BibitemOpen
  \bibfield  {author} {\bibinfo {author} {\bibfnamefont {P.~T.}\ \bibnamefont
  {Dumitrescu}}, \bibinfo {author} {\bibfnamefont {A.}~\bibnamefont
  {Goremykina}}, \bibinfo {author} {\bibfnamefont {S.~A.}\ \bibnamefont
  {Parameswaran}}, \bibinfo {author} {\bibfnamefont {M.}~\bibnamefont
  {Serbyn}}, \ and\ \bibinfo {author} {\bibfnamefont {R.}~\bibnamefont
  {Vasseur}},\ }\href {\doibase 10.1103/PhysRevB.99.094205} {\bibfield
  {journal} {\bibinfo  {journal} {Phys. Rev. B}\ }\textbf {\bibinfo {volume}
  {99}},\ \bibinfo {pages} {094205} (\bibinfo {year} {2019})}\BibitemShut
  {NoStop}%
\bibitem [{\citenamefont {Lev}\ \emph {et~al.}(2017)\citenamefont {Lev},
  \citenamefont {Kennes}, \citenamefont {Kl\"ockner}, \citenamefont
  {Reichman},\ and\ \citenamefont {Karrasch}}]{Bar_Lev_2017}%
  \BibitemOpen
  \bibfield  {author} {\bibinfo {author} {\bibfnamefont {Y.~B.}\ \bibnamefont
  {Lev}}, \bibinfo {author} {\bibfnamefont {D.~M.}\ \bibnamefont {Kennes}},
  \bibinfo {author} {\bibfnamefont {C.}~\bibnamefont {Kl\"ockner}}, \bibinfo
  {author} {\bibfnamefont {D.~R.}\ \bibnamefont {Reichman}}, \ and\ \bibinfo
  {author} {\bibfnamefont {C.}~\bibnamefont {Karrasch}},\ }\href {\doibase
  10.1209/0295-5075/119/37003} {\bibfield  {journal} {\bibinfo  {journal}
  {{EPL} (Europhysics Letters)}\ }\textbf {\bibinfo {volume} {119}},\ \bibinfo
  {pages} {37003} (\bibinfo {year} {2017})}\BibitemShut {NoStop}%
\bibitem [{\citenamefont {Chayes}\ \emph {et~al.}(1986)\citenamefont {Chayes},
  \citenamefont {Chayes}, \citenamefont {Fisher},\ and\ \citenamefont
  {Spencer}}]{Chayes1986}%
  \BibitemOpen
  \bibfield  {author} {\bibinfo {author} {\bibfnamefont {J.~T.}\ \bibnamefont
  {Chayes}}, \bibinfo {author} {\bibfnamefont {L.}~\bibnamefont {Chayes}},
  \bibinfo {author} {\bibfnamefont {D.~S.}\ \bibnamefont {Fisher}}, \ and\
  \bibinfo {author} {\bibfnamefont {T.}~\bibnamefont {Spencer}},\ }\href
  {\doibase 10.1103/PhysRevLett.57.2999} {\bibfield  {journal} {\bibinfo
  {journal} {Phys. Rev. Lett.}\ }\textbf {\bibinfo {volume} {57}},\ \bibinfo
  {pages} {2999} (\bibinfo {year} {1986})}\BibitemShut {NoStop}%
\bibitem [{Note3()}]{Note3}%
  \BibitemOpen
  \bibinfo {note} {\label {fn23} When working with quasi-periodic potential
  (Aubry-Andr\'e model) we choose, $\epsilon _x {=} W \protect \qopname \relax
  o{cos}(2\pi \alpha x + \phi )$ with $\alpha =2/(\protect \sqrt {5}+1)$ and
  $\phi $ is the random phase distributed uniformly between $[0, 2 \pi
  ]$}\BibitemShut {NoStop}%
\bibitem [{Note4()}]{Note4}%
  \BibitemOpen
  \bibinfo {note} {Here, we have used that in the infinite temperature limit
  the density is homogeneous and equal to $1/2$.}\BibitemShut {Stop}%
\bibitem [{\citenamefont {Wei\ss{}e}\ \emph {et~al.}(2006)\citenamefont
  {Wei\ss{}e}, \citenamefont {Wellein}, \citenamefont {Alvermann},\ and\
  \citenamefont {Fehske}}]{Wei06}%
  \BibitemOpen
  \bibfield  {author} {\bibinfo {author} {\bibfnamefont {A.}~\bibnamefont
  {Wei\ss{}e}}, \bibinfo {author} {\bibfnamefont {G.}~\bibnamefont {Wellein}},
  \bibinfo {author} {\bibfnamefont {A.}~\bibnamefont {Alvermann}}, \ and\
  \bibinfo {author} {\bibfnamefont {H.}~\bibnamefont {Fehske}},\ }\href
  {\doibase 10.1103/RevModPhys.78.275} {\bibfield  {journal} {\bibinfo
  {journal} {Rev. Mod. Phys.}\ }\textbf {\bibinfo {volume} {78}},\ \bibinfo
  {pages} {275} (\bibinfo {year} {2006})}\BibitemShut {NoStop}%
\bibitem [{Note5()}]{Note5}%
  \BibitemOpen
  \bibinfo {note} {Alternative possibilities could be (i) $\protect \mathrm
  {b}_1(W_{\protect \text {c}_1}){=}0$ or (ii) $\protect \mathrm
  {b}_1(W_{\protect \text {c}_1}) {\to }\infty $. However, in our scenario we
  do not follow these directions.}\BibitemShut {Stop}%
\bibitem [{\citenamefont {Khemani}\ \emph
  {et~al.}(2017{\natexlab{b}})\citenamefont {Khemani}, \citenamefont {Lim},
  \citenamefont {Sheng},\ and\ \citenamefont {Huse}}]{Khemani2017}%
  \BibitemOpen
  \bibfield  {author} {\bibinfo {author} {\bibfnamefont {V.}~\bibnamefont
  {Khemani}}, \bibinfo {author} {\bibfnamefont {S.~P.}\ \bibnamefont {Lim}},
  \bibinfo {author} {\bibfnamefont {D.~N.}\ \bibnamefont {Sheng}}, \ and\
  \bibinfo {author} {\bibfnamefont {D.~A.}\ \bibnamefont {Huse}},\ }\href
  {\doibase 10.1103/PhysRevX.7.021013} {\bibfield  {journal} {\bibinfo
  {journal} {Phys. Rev. X}\ }\textbf {\bibinfo {volume} {7}},\ \bibinfo {pages}
  {021013} (\bibinfo {year} {2017}{\natexlab{b}})}\BibitemShut {NoStop}%
\bibitem [{\citenamefont {Gray}\ \emph {et~al.}(2018)\citenamefont {Gray},
  \citenamefont {Bose},\ and\ \citenamefont {Bayat}}]{Gray2018}%
  \BibitemOpen
  \bibfield  {author} {\bibinfo {author} {\bibfnamefont {J.}~\bibnamefont
  {Gray}}, \bibinfo {author} {\bibfnamefont {S.}~\bibnamefont {Bose}}, \ and\
  \bibinfo {author} {\bibfnamefont {A.}~\bibnamefont {Bayat}},\ }\href
  {\doibase 10.1103/PhysRevB.97.201105} {\bibfield  {journal} {\bibinfo
  {journal} {Phys. Rev. B}\ }\textbf {\bibinfo {volume} {97}},\ \bibinfo
  {pages} {201105} (\bibinfo {year} {2018})}\BibitemShut {NoStop}%
\bibitem [{\citenamefont {Lenar\ifmmode \check{c}\else
  \v{c}\fi{}i\ifmmode~\check{c}\else \v{c}\fi{}}\ \emph
  {et~al.}(2018)\citenamefont {Lenar\ifmmode \check{c}\else
  \v{c}\fi{}i\ifmmode~\check{c}\else \v{c}\fi{}}, \citenamefont {Altman},\ and\
  \citenamefont {Rosch}}]{Lenarcic2018}%
  \BibitemOpen
  \bibfield  {author} {\bibinfo {author} {\bibfnamefont {Z.}~\bibnamefont
  {Lenar\ifmmode \check{c}\else \v{c}\fi{}i\ifmmode~\check{c}\else
  \v{c}\fi{}}}, \bibinfo {author} {\bibfnamefont {E.}~\bibnamefont {Altman}}, \
  and\ \bibinfo {author} {\bibfnamefont {A.}~\bibnamefont {Rosch}},\ }\href
  {\doibase 10.1103/PhysRevLett.121.267603} {\bibfield  {journal} {\bibinfo
  {journal} {Phys. Rev. Lett.}\ }\textbf {\bibinfo {volume} {121}},\ \bibinfo
  {pages} {267603} (\bibinfo {year} {2018})}\BibitemShut {NoStop}%
\bibitem [{\citenamefont {{Morningstar}}\ and\ \citenamefont
  {{Huse}}(2019)}]{Morningstar2019}%
  \BibitemOpen
  \bibfield  {author} {\bibinfo {author} {\bibfnamefont {A.}~\bibnamefont
  {{Morningstar}}}\ and\ \bibinfo {author} {\bibfnamefont {D.~A.}\ \bibnamefont
  {{Huse}}},\ }\href@noop {} {\bibfield  {journal} {\bibinfo  {journal} {arXiv
  e-prints}\ ,\ \bibinfo {eid} {arXiv:1903.02001}} (\bibinfo {year} {2019})},\
  \Eprint {http://arxiv.org/abs/1903.02001} {arXiv:1903.02001
  [cond-mat.stat-mech]} \BibitemShut {NoStop}%
\bibitem [{\citenamefont {Goremykina}\ \emph {et~al.}(2019)\citenamefont
  {Goremykina}, \citenamefont {Vasseur},\ and\ \citenamefont
  {Serbyn}}]{Goremykina2018}%
  \BibitemOpen
  \bibfield  {author} {\bibinfo {author} {\bibfnamefont {A.}~\bibnamefont
  {Goremykina}}, \bibinfo {author} {\bibfnamefont {R.}~\bibnamefont {Vasseur}},
  \ and\ \bibinfo {author} {\bibfnamefont {M.}~\bibnamefont {Serbyn}},\ }\href
  {\doibase 10.1103/PhysRevLett.122.040601} {\bibfield  {journal} {\bibinfo
  {journal} {Phys. Rev. Lett.}\ }\textbf {\bibinfo {volume} {122}},\ \bibinfo
  {pages} {040601} (\bibinfo {year} {2019})}\BibitemShut {NoStop}%
\bibitem [{\citenamefont {Zhang}\ \emph {et~al.}(2016)\citenamefont {Zhang},
  \citenamefont {Zhao}, \citenamefont {Devakul},\ and\ \citenamefont
  {Huse}}]{Zhang2016}%
  \BibitemOpen
  \bibfield  {author} {\bibinfo {author} {\bibfnamefont {L.}~\bibnamefont
  {Zhang}}, \bibinfo {author} {\bibfnamefont {B.}~\bibnamefont {Zhao}},
  \bibinfo {author} {\bibfnamefont {T.}~\bibnamefont {Devakul}}, \ and\
  \bibinfo {author} {\bibfnamefont {D.~A.}\ \bibnamefont {Huse}},\ }\href
  {\doibase 10.1103/PhysRevB.93.224201} {\bibfield  {journal} {\bibinfo
  {journal} {Phys. Rev. B}\ }\textbf {\bibinfo {volume} {93}},\ \bibinfo
  {pages} {224201} (\bibinfo {year} {2016})}\BibitemShut {NoStop}%
\bibitem [{\citenamefont {Thiery}\ \emph {et~al.}(2018)\citenamefont {Thiery},
  \citenamefont {Huveneers}, \citenamefont {M\"uller},\ and\ \citenamefont
  {De~Roeck}}]{Thiery2017}%
  \BibitemOpen
  \bibfield  {author} {\bibinfo {author} {\bibfnamefont {T.}~\bibnamefont
  {Thiery}}, \bibinfo {author} {\bibfnamefont {F.}~\bibnamefont {Huveneers}},
  \bibinfo {author} {\bibfnamefont {M.}~\bibnamefont {M\"uller}}, \ and\
  \bibinfo {author} {\bibfnamefont {W.}~\bibnamefont {De~Roeck}},\ }\href
  {\doibase 10.1103/PhysRevLett.121.140601} {\bibfield  {journal} {\bibinfo
  {journal} {Phys. Rev. Lett.}\ }\textbf {\bibinfo {volume} {121}},\ \bibinfo
  {pages} {140601} (\bibinfo {year} {2018})}\BibitemShut {NoStop}%
\bibitem [{\citenamefont {{Herviou}}\ \emph {et~al.}(2018)\citenamefont
  {{Herviou}}, \citenamefont {{Bera}},\ and\ \citenamefont
  {{Bardarson}}}]{Loic2018}%
  \BibitemOpen
  \bibfield  {author} {\bibinfo {author} {\bibfnamefont {L.}~\bibnamefont
  {{Herviou}}}, \bibinfo {author} {\bibfnamefont {S.}~\bibnamefont {{Bera}}}, \
  and\ \bibinfo {author} {\bibfnamefont {J.~H.}\ \bibnamefont {{Bardarson}}},\
  }\href@noop {} {\bibfield  {journal} {\bibinfo  {journal} {arXiv e-prints}\
  ,\ \bibinfo {eid} {arXiv:1811.01925}} (\bibinfo {year} {2018})},\ \Eprint
  {http://arxiv.org/abs/1811.01925} {arXiv:1811.01925 [cond-mat.dis-nn]}
  \BibitemShut {NoStop}%
\bibitem [{Note6()}]{Note6}%
  \BibitemOpen
  \bibinfo {note} {With respect to $W_\protect \text {c}$ a similar conclusion
  can also be drawn from the scaling of the Schmidt-gap displayed in Fig. 5 of
  the same work. \cite {Doggen2018}}\BibitemShut {NoStop}%
\bibitem [{\citenamefont {Prelov\ifmmode~\check{s}\else \v{s}\fi{}ek}\ and\
  \citenamefont {Herbrych}(2017)}]{PrelovsekPRB17}%
  \BibitemOpen
  \bibfield  {author} {\bibinfo {author} {\bibfnamefont {P.}~\bibnamefont
  {Prelov\ifmmode~\check{s}\else \v{s}\fi{}ek}}\ and\ \bibinfo {author}
  {\bibfnamefont {J.}~\bibnamefont {Herbrych}},\ }\href {\doibase
  10.1103/PhysRevB.96.035130} {\bibfield  {journal} {\bibinfo  {journal} {Phys.
  Rev. B}\ }\textbf {\bibinfo {volume} {96}},\ \bibinfo {pages} {035130}
  (\bibinfo {year} {2017})}\BibitemShut {NoStop}%
\bibitem [{\citenamefont {Mierzejewski}\ \emph {et~al.}(2016)\citenamefont
  {Mierzejewski}, \citenamefont {Herbrych},\ and\ \citenamefont
  {Prelov\ifmmode~\check{s}\else \v{s}\fi{}ek}}]{MierzejewskiPRB16}%
  \BibitemOpen
  \bibfield  {author} {\bibinfo {author} {\bibfnamefont {M.}~\bibnamefont
  {Mierzejewski}}, \bibinfo {author} {\bibfnamefont {J.}~\bibnamefont
  {Herbrych}}, \ and\ \bibinfo {author} {\bibfnamefont {P.}~\bibnamefont
  {Prelov\ifmmode~\check{s}\else \v{s}\fi{}ek}},\ }\href {\doibase
  10.1103/PhysRevB.94.224207} {\bibfield  {journal} {\bibinfo  {journal} {Phys.
  Rev. B}\ }\textbf {\bibinfo {volume} {94}},\ \bibinfo {pages} {224207}
  (\bibinfo {year} {2016})}\BibitemShut {NoStop}%
\bibitem [{\citenamefont {Atas}\ and\ \citenamefont
  {Bogomolny}(2014)}]{Bogomolny2014}%
  \BibitemOpen
  \bibfield  {author} {\bibinfo {author} {\bibfnamefont {Y.~Y.}\ \bibnamefont
  {Atas}}\ and\ \bibinfo {author} {\bibfnamefont {E.}~\bibnamefont
  {Bogomolny}},\ }\href {\doibase 10.1098/rsta.2012.0520} {\bibfield  {journal}
  {\bibinfo  {journal} {Philosophical Transactions of the Royal Society A:
  Mathematical, Physical and Engineering Sciences}\ }\textbf {\bibinfo {volume}
  {372}},\ \bibinfo {pages} {20120520} (\bibinfo {year} {2014})}\BibitemShut
  {NoStop}%
\bibitem [{\citenamefont {Gornyi}\ \emph {et~al.}(2017)\citenamefont {Gornyi},
  \citenamefont {Mirlin}, \citenamefont {Polyakov},\ and\ \citenamefont
  {Burin}}]{Gornyi2017}%
  \BibitemOpen
  \bibfield  {author} {\bibinfo {author} {\bibfnamefont {I.~V.}\ \bibnamefont
  {Gornyi}}, \bibinfo {author} {\bibfnamefont {A.~D.}\ \bibnamefont {Mirlin}},
  \bibinfo {author} {\bibfnamefont {D.~G.}\ \bibnamefont {Polyakov}}, \ and\
  \bibinfo {author} {\bibfnamefont {A.~L.}\ \bibnamefont {Burin}},\ }\href
  {\doibase 10.1002/andp.201600360} {\bibfield  {journal} {\bibinfo  {journal}
  {Annalen der Physik}\ }\textbf {\bibinfo {volume} {529}},\ \bibinfo {pages}
  {1600360} (\bibinfo {year} {2017})}\BibitemShut {NoStop}%
\bibitem [{\citenamefont {{Mac{\'e}}}\ \emph {et~al.}(2018)\citenamefont
  {{Mac{\'e}}}, \citenamefont {{Alet}},\ and\ \citenamefont
  {{Laflorencie}}}]{MaceMultifractality2018}%
  \BibitemOpen
  \bibfield  {author} {\bibinfo {author} {\bibfnamefont {N.}~\bibnamefont
  {{Mac{\'e}}}}, \bibinfo {author} {\bibfnamefont {F.}~\bibnamefont {{Alet}}},
  \ and\ \bibinfo {author} {\bibfnamefont {N.}~\bibnamefont {{Laflorencie}}},\
  }\href@noop {} {\bibfield  {journal} {\bibinfo  {journal} {arXiv e-prints}\
  ,\ \bibinfo {eid} {arXiv:1812.10283}} (\bibinfo {year} {2018})},\ \Eprint
  {http://arxiv.org/abs/1812.10283} {arXiv:1812.10283 [cond-mat.dis-nn]}
  \BibitemShut {NoStop}%
\bibitem [{\citenamefont {Serbyn}\ \emph {et~al.}(2015)\citenamefont {Serbyn},
  \citenamefont {Papi\ifmmode~\acute{c}\else \'{c}\fi{}},\ and\ \citenamefont
  {Abanin}}]{Serbyn2015}%
  \BibitemOpen
  \bibfield  {author} {\bibinfo {author} {\bibfnamefont {M.}~\bibnamefont
  {Serbyn}}, \bibinfo {author} {\bibfnamefont {Z.}~\bibnamefont
  {Papi\ifmmode~\acute{c}\else \'{c}\fi{}}}, \ and\ \bibinfo {author}
  {\bibfnamefont {D.~A.}\ \bibnamefont {Abanin}},\ }\href {\doibase
  10.1103/PhysRevX.5.041047} {\bibfield  {journal} {\bibinfo  {journal} {Phys.
  Rev. X}\ }\textbf {\bibinfo {volume} {5}},\ \bibinfo {pages} {041047}
  (\bibinfo {year} {2015})}\BibitemShut {NoStop}%
\bibitem [{Note7()}]{Note7}%
  \BibitemOpen
  \bibinfo {note} {We express our gratitude to Ehud Altman for drawing our
  attention to this work.}\BibitemShut {Stop}%
\bibitem [{\citenamefont {Bertrand}\ and\ \citenamefont
  {Garc\'{\i}a-Garc\'{\i}a}(2016)}]{Bertrand2016}%
  \BibitemOpen
  \bibfield  {author} {\bibinfo {author} {\bibfnamefont {C.~L.}\ \bibnamefont
  {Bertrand}}\ and\ \bibinfo {author} {\bibfnamefont {A.~M.}\ \bibnamefont
  {Garc\'{\i}a-Garc\'{\i}a}},\ }\href {\doibase 10.1103/PhysRevB.94.144201}
  {\bibfield  {journal} {\bibinfo  {journal} {Phys. Rev. B}\ }\textbf {\bibinfo
  {volume} {94}},\ \bibinfo {pages} {144201} (\bibinfo {year}
  {2016})}\BibitemShut {NoStop}%
\bibitem [{\citenamefont {Sierant}\ and\ \citenamefont
  {Zakrzewski}(2019)}]{SierantPRB19}%
  \BibitemOpen
  \bibfield  {author} {\bibinfo {author} {\bibfnamefont {P.}~\bibnamefont
  {Sierant}}\ and\ \bibinfo {author} {\bibfnamefont {J.}~\bibnamefont
  {Zakrzewski}},\ }\href {\doibase 10.1103/PhysRevB.99.104205} {\bibfield
  {journal} {\bibinfo  {journal} {Phys. Rev. B}\ }\textbf {\bibinfo {volume}
  {99}},\ \bibinfo {pages} {104205} (\bibinfo {year} {2019})}\BibitemShut
  {NoStop}%
\bibitem [{\citenamefont {Tikhonov}\ \emph {et~al.}(2016)\citenamefont
  {Tikhonov}, \citenamefont {Mirlin},\ and\ \citenamefont
  {Skvortsov}}]{TikhonovRRG16}%
  \BibitemOpen
  \bibfield  {author} {\bibinfo {author} {\bibfnamefont {K.~S.}\ \bibnamefont
  {Tikhonov}}, \bibinfo {author} {\bibfnamefont {A.~D.}\ \bibnamefont
  {Mirlin}}, \ and\ \bibinfo {author} {\bibfnamefont {M.~A.}\ \bibnamefont
  {Skvortsov}},\ }\href {\doibase 10.1103/PhysRevB.94.220203} {\bibfield
  {journal} {\bibinfo  {journal} {Phys. Rev. B}\ }\textbf {\bibinfo {volume}
  {94}},\ \bibinfo {pages} {220203} (\bibinfo {year} {2016})}\BibitemShut
  {NoStop}%
\bibitem [{\citenamefont {Serbyn}\ \emph {et~al.}(2013)\citenamefont {Serbyn},
  \citenamefont {Papi\ifmmode~\acute{c}\else \'{c}\fi{}},\ and\ \citenamefont
  {Abanin}}]{Serbyn2013}%
  \BibitemOpen
  \bibfield  {author} {\bibinfo {author} {\bibfnamefont {M.}~\bibnamefont
  {Serbyn}}, \bibinfo {author} {\bibfnamefont {Z.}~\bibnamefont
  {Papi\ifmmode~\acute{c}\else \'{c}\fi{}}}, \ and\ \bibinfo {author}
  {\bibfnamefont {D.~A.}\ \bibnamefont {Abanin}},\ }\href {\doibase
  10.1103/PhysRevLett.111.127201} {\bibfield  {journal} {\bibinfo  {journal}
  {Phys. Rev. Lett.}\ }\textbf {\bibinfo {volume} {111}},\ \bibinfo {pages}
  {127201} (\bibinfo {year} {2013})}\BibitemShut {NoStop}%
\bibitem [{\citenamefont {Huse}\ \emph {et~al.}(2014)\citenamefont {Huse},
  \citenamefont {Nandkishore},\ and\ \citenamefont {Oganesyan}}]{Huse2014}%
  \BibitemOpen
  \bibfield  {author} {\bibinfo {author} {\bibfnamefont {D.~A.}\ \bibnamefont
  {Huse}}, \bibinfo {author} {\bibfnamefont {R.}~\bibnamefont {Nandkishore}}, \
  and\ \bibinfo {author} {\bibfnamefont {V.}~\bibnamefont {Oganesyan}},\ }\href
  {\doibase 10.1103/PhysRevB.90.174202} {\bibfield  {journal} {\bibinfo
  {journal} {Phys. Rev. B}\ }\textbf {\bibinfo {volume} {90}},\ \bibinfo
  {pages} {174202} (\bibinfo {year} {2014})}\BibitemShut {NoStop}%
\bibitem [{Note8()}]{Note8}%
  \BibitemOpen
  \bibinfo {note} {When working with quasi-periodic potential (Aubry-Andr\'e
  model) we choose, $\epsilon _x {=} W \protect \qopname \relax o{cos}(2\pi
  \alpha x + \phi )$ with $\alpha =2/(\protect \sqrt {5}+1)$ and $\phi $ is the
  random phase distributed uniformly between $[0, 2 \pi ]$}\BibitemShut
  {NoStop}%
\bibitem [{\citenamefont {Luitz}\ \emph {et~al.}(2016)\citenamefont {Luitz},
  \citenamefont {Laflorencie},\ and\ \citenamefont {Alet}}]{Luitz2016}%
  \BibitemOpen
  \bibfield  {author} {\bibinfo {author} {\bibfnamefont {D.~J.}\ \bibnamefont
  {Luitz}}, \bibinfo {author} {\bibfnamefont {N.}~\bibnamefont {Laflorencie}},
  \ and\ \bibinfo {author} {\bibfnamefont {F.}~\bibnamefont {Alet}},\ }\href
  {\doibase 10.1103/PhysRevB.93.060201} {\bibfield  {journal} {\bibinfo
  {journal} {Phys. Rev. B}\ }\textbf {\bibinfo {volume} {93}},\ \bibinfo
  {pages} {060201} (\bibinfo {year} {2016})}\BibitemShut {NoStop}%
\end{thebibliography}%

\end{document}